\documentclass{jkas}

\usepackage{color}

\def\year{2020} 
\def\month{XXX} 
\def\volume{00} 
\def\issue{0} 
\def\beginpage{1} 
\def\endpage{99} 
\def\received{Xxxxx 00, 2020} 
\def\accepted{Xxxxx 00, 2020} 
\date{Received \received; accepted \accepted}

\newcommand{\oiii}{[\mbox{O\,\textsc{iii}}]}
\newcommand{\nii}{[\mbox{N\,\textsc{ii}}]}
\newcommand{\sii}{[\mbox{S\,\textsc{ii}}]}
\newcommand{\ha}{H$\alpha$}   
\newcommand{\hb}{H$\beta$}  

\newcommand{\kms}{km s$^{-1}$}
\newcommand{\msigma}{M$_{\rm BH}-\sigma_{*}$}
\newcommand{\msun}{M$_{\odot}$}

\newcommand{\SVD}{$\sigma_{*}$}
\newcommand{\ergs}{erg s$^{-1}$} 
\newcommand{\mbh}{M$_{\rm BH}$}

\newcommand{\arcsec}{\hbox{$^{\prime\prime}$}}

\title{Spatially Resolved Kinematics of gas and stars in hidden type 1 AGNs}

\author[1,2]{Donghoon Son}
\author[1,3]{Jong-Hak Woo}
\author[1]{Da-In Eun}
\author[1]{Hojin Cho}
\author[4]{Marios Karouzos}
\author[1]{Songyeon Park}

\affil[1]{Astronomy Program, Department of Physics and Astronomy, Seoul National University, Seoul 08826, Republic of Korea}
\affil[2]{Department of Astronomy and Center for Galaxy Evolution Research, Yonsei University, Seoul 03722, Republic of Korea}
\affil[3]{Korea Astronomy and Space Science Institute, Daejeon 34055, Republic of Korea}
\affil[4]{Nature Astronomy, Springer Nature, 4 Crinan Street, N1 9XW, London, UK}

\def\corrauthor{Jong-Hak Woo (woo@astro.snu.ac.kr)}

\def\runningauthor{Donghoon Son}
\def\runningtitle{Kinematics of gas and stars in type 1 AGN}

\def\keywords{galaxies: active --- galaxies: kinematics and dynamics --- quasars: emission lines}

\def\abstracttext{
We present the spatially resolved gas and stellar kinematics of a sample of ten hidden type 1 AGNs in order to investigate the 
true nature of the central source and the scaling relation with host galaxy stellar velocity dispersion. The sample is selected 
from a large number of hidden type 1 AGN, which are identified based on the presence of a broad component in the \ha\ line profile (i.e., full-width-at-half-maximum $>$ $\sim$1000 \kms), while they are often mis-classified as type 2 AGN because AGN continuum and broad emission lines are weak or obscured in the optical spectral range. 
We used the Blue Channel Spectrograph at the 6.5-m MMT (Multiple Mirror Telescope) to obtain long-slit data.
We detected a broad \hb\ for only two targets, however, the presence of a strong broad \ha\ indicates that these AGNs are low-luminosity type 1 AGNs.
We measured the velocity, velocity dispersion and flux of stellar continuum and gas emission lines (i.e., \hb\ and \oiii) as a function of distance from the center with a spatial scale of 0.3 arcsec pixel$^{-1}$.
Spatially resolved gas kinematics traced by \hb\ or \oiii\ are generally similar to stellar kinematics except for the very center, 
where signatures of gas outflows are detected. 
We compare the luminosity-weighted effective stellar velocity dispersion with black hole mass, finding that these hidden type 1 AGN
with relatively low back hole mass follow the scaling relation of the reverberation-mapped type 1 AGN and more massive inactive galaxies. }

\begin{document}

\jkashead 

\section{Introduction}
The black hole mass (M$_{\rm BH}$) correlation with stellar velocity dispersion ($\sigma_*$) may imply a connection between galaxy evolution and black hole growth \citep[e.g.,][]{KH13,Woo+13}.
In the present day, both inactive and active galaxies seem to follow the same \msigma\ relation regardless of the black hole activity 
\citep{Park+12,Woo+13,Woo+15}. However, it is unclear whether the correlation 
extends to lower mass and low-luminosity regime, down to intermediate-mass black holes \citep[e.g.][]{Woo+19, Greene+19}

Type 1 AGNs have been generally used to investigate the correlation between black hole mass and host galaxy properties 
since black hole mass can be relatively easily determined based on single-epoch spectrum for AGN with broad emission lines, using various mass estimators \citep[e.g.,][]{Kaspi+00, Kaspi+05, McGill+08, Bentz+09, Bentz+13, Park+15, Woo+15, Woo+18}. However, for high-luminosity AGN, host galaxy properties, i.e., stellar velocity dispersion, bulge luminosity, and stellar mass, are
difficult to measure since the strong continuum source at the galactic nucleus over-shines an entire host galaxy. 
On the other hand, type 2 AGNs are easy to determine their host galaxy properties due to the lack of the strong continuum in the observed spectrum. 
In turn, their black hole mass is difficult to determine as the gas in the broad line region (BLR) is obscured by a dusty torus. 

Between type 1 and type 2 AGNs, there is an interesting class of AGN, which present broad emission lines, but continuum is dominated by the stellar component in the host galaxy. Type 2 AGNs are often classified among emission-line galaxies based on the narrow emission line flux ratio diagnosis \citep{bpt}. However, a small fraction of these type 2 AGNs presents a broad component in the \ha\ emission line, hence, these objects are qualified as type 1 AGN. In fact many SDSS type 2 AGNs, which are identified based on the emission line flux ratios, are previously known or re-classified by follow up studies as type 1 AGN. Since the continuum of these AGNs is dominated by stellar component, and the broad component is often
not detected, we will broadly call them hidden type 1 AGNs.

In our previous study, we searched for hidden type 1 AGN among the local type 2 AGN sample, which were investigated based on the Sloan Digital Sky Survey Data Release 7 \citep{Woo+14}.
Among 4,113 local type 2 AGNs at $0.02 < z < 0.05$, \citet{Woo+14} found a sample of 142 type 1 AGNs by detecting a broad component in \ha, 
of which the Full-Width at Half-Maximum (FWHM) of the line profile ranges from 1700 to $\sim$20,000 \kms\ 
based on the spectral decomposition analysis including stellar population models and Gaussian modeling of emission line components. 
The fraction of the hidden type 1 AGN among type 2 AGN sample is $\sim$3.5\%, implying that a large number of missing type 1 AGN population may exist \citep{Woo+14}. 
Similarly, \citet{Oh+15} identified 1,835 type 1 AGNs  at $z < 0.2$, which show a weak broad emission line. 
As a follow-up study, \citet{Eun+17} carefully investigated $\sim$24,000 type 2 AGNs at $z < 0.1$, which were selected from SDSS DR7 sample defined by \citet{BW14}, and found a sample of 611 hidden type 1 AGNs with broad \ha\ lines with FWHM $>$ 1000 \kms.

These hidden type 1 AGNs are valuable targets for further studying the black hole - galaxy coevolution,
since they are likely to be low-luminosity AGN with relatively weak continuum (the median
Eddington ratio is 1\%). Thus, their host galaxy properties can be easily studied while the
mass of the central black hole can be estimated from the broad component of \ha\
using various single-epoch mass estimators \citep[e.g.,][]{WU02, Park+12b, Bentz+13,Woo+15}.

Although dynamical black hole mass measurements have been improved over the last decade to better define
the local \msigma\ relation \citep[see][]{KH13}, not as much attention has been paid to stellar 
velocity dispersion (SVD).
Currently, there are two limitations in the SVD measurements, 
which arguably lead to systematic uncertainties in the local \msigma\ relation, particularly at the low mass regime. 
First, although SVD has been measured from spatially resolved stellar kinematics for most early-type
galaxies based on high quality data  \citep[e.g.,][]{MM13, Kang+13}, the SVDs of a number of
late-type galaxies have been collected from old literature. These values were often measured 
from single-aperture spectra, which were extracted with different aperture sizes. 
The lack of spatially resolved measurements, 
and inhomogeneous data quality and analysis of these late-type galaxies increase the systematic uncertainties
of the \msigma\ relation on the lower mass regime, and prevents us from properly constraining the intrinsic scatter of the \msigma\ relation. 
Second, rotation and inclination effects of the stellar disk have not been carefully accounted in measuring SVD.
From the spatially resolved stellar kinematics, the effective SVD 
has been calculated by integrating the luminosity-weighted sum of velocity dispersion and velocity (hereafter, rotation-added ($RA$)) as
\begin{equation}
\sigma^{2}_{*,RA} = { \int^{R_e}_{-R_e}(\sigma_{*}(r)^{2}+V(r)^{2})I(r)dr \over \int^{R_e}_{-R_e}I(r)dr },
 \label{eq_RA}
\end{equation}
out to the effective radius ($R_{e}$), for defining the \msigma\ relation \citep{Gul+09, MM13, KH13}. 
However, this definition of SVD is not a pure velocity dispersion since it includes a contribution from the disk rotation. 
If black hole mass correlates only with bulge properties, but not with disk properties, 
then we may need to use a SVD corresponding to the bulge potential without a disk contribution. 
Depending on adding or excluding the disk contribution, 
the effective SVD may make substantial difference for late-type galaxies, hence in deriving the \msigma\ relation.
For example, \cite{Ja+11} showed that the SVD of a Sa galaxy, NGC 4594
changes from 297 \kms\ to 200 \kms\ if the rotation is excluded in calculating the effective SVD (see their Fig. 9).
\cite{Bell+14} investigated the effect of rotational broadening if spatially resolved data are not available. 
Thus, it is important to explore the \msigma\ relation at the lower mass regime, with spatially resolved stellar kinematics. \

In this study, we present a study of spatially resolved kinematics of gas and stars using a sample of 10 hidden type 1 AGNs,
in order to investigate the true nature of the central source (i.e., type 1 or type 2) and the black hole mass correlation with 
stellar velocity dispersion. We describe the sample selection, observation and data reduction in Section 2.
In Section~\ref{measure}, we describe analysis of gas and stellar kinematics, which are followed by the results in Section~\ref{result}.
Section~\ref{summary} presents a summary and conclusions. Throughout the paper, we used cosmological parameters, $H_0 = 70$ \kms~ Mpc$^{-1}$, $\Omega_{m} = 0.30$, and $\Omega_{\Lambda} = 0.70$.


\begin{figure*}[htp]
\centering
\includegraphics[width=\textwidth]{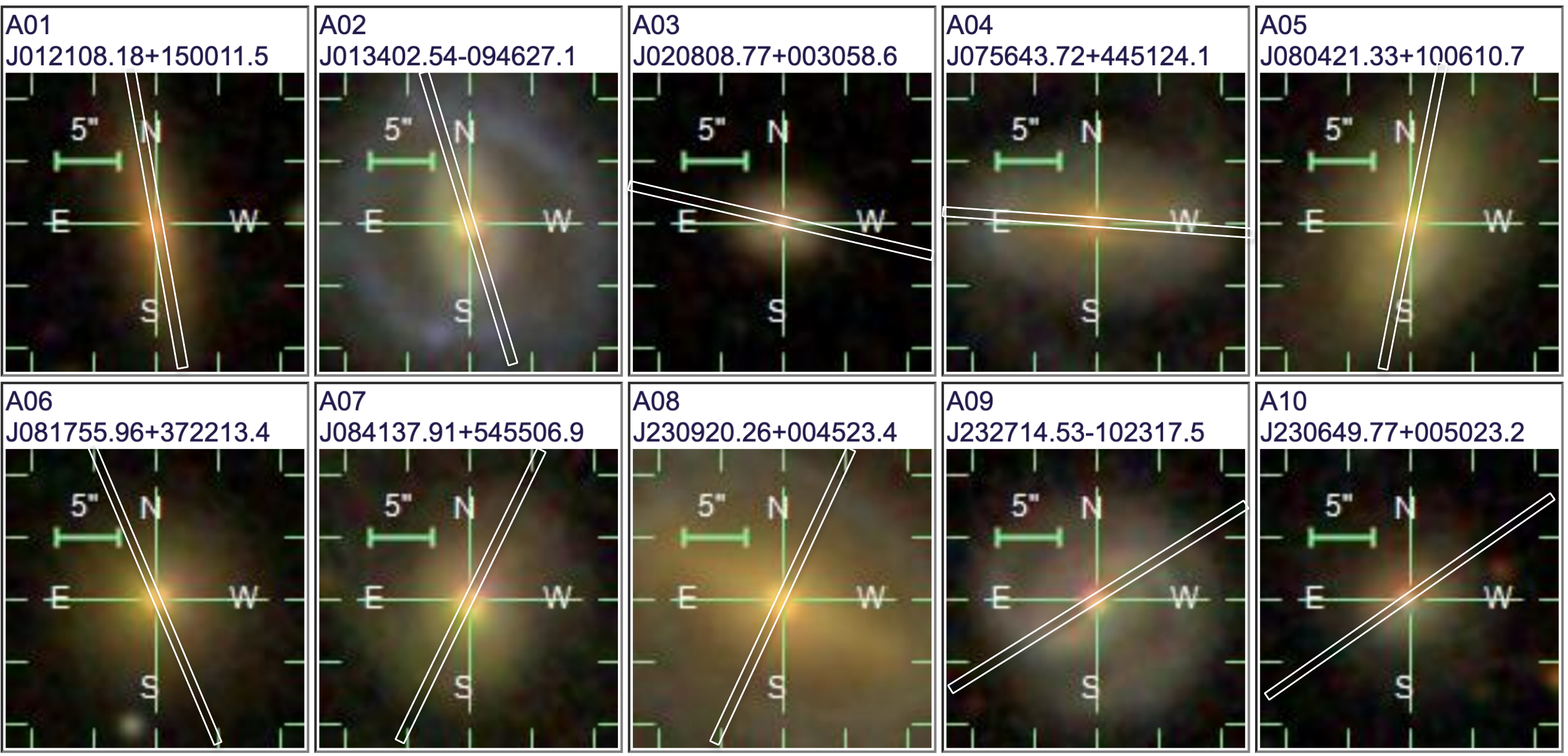}
\caption{The SDSS {\it gri} composite images provided by the SDSS SkyServer. The position angle of the slit is represented with an empty bar for each target. }
\label{slit}
\end{figure*}

\section{Observations}

\subsection{Sample selection\label{sample}}

We selected a pilot sample of ten type 1 AGNs from our previous studies by \citet{Eun+17}, who identified a sample of 611 hidden-type 1 AGNs out to $z=0.1$,
by detecting a broad components in the \ha\ line. Note that six out of ten AGNs were correctly classified as type 1 AGNs by \citet{VV06}.
However, as the continuum of these targets is dominated by stellar component because of the lack of AGN continuum, and the broad component in \hb\ is not detected,
they are also useful for stellar velocity dispersion measurements. 
While this sample is small and not representative for all hidden type 1 AGNs, we selected the sample with 
several criteria. First, we selected targets with SDSS-based SVD larger than 90 \kms\ 
to avoid uncertainty of velocity dispersion measurements due to a limited spectral resolution. 
Second, we selected observable targets (i.e., $23h < R.A. < 09h$),
which were optimal for the scheduled night in 2015B. 


\begin{table*}[htp]
\centering
\caption{Target properties and log of observation \label{target}}
\begin{tabular}{cccccccccrcc}
\toprule
ID & SDSS Name & z & g & b/a & FWHM$_{\rm H\alpha}$ & log L$_{\rm H\alpha}$  & \SVD & log M$_{\rm BH}$ & PA & EXPT & SN \\
 & & & & & (\kms) & (\ergs) & (\kms) & (M$_\odot$) & (deg) & (min) & \\ 
(1) & (2) & (3) & (4) & (5) & (6) & (7) & (8) & (9) & (10) & (11) & (12) \\
\midrule
A01 & J012108.17+150011.5    & 0.054 & 17.1 & 0.41 & 3311 & 40.8 & 180.4 & 7.12 & 10       & 40 & 17 \\	
A02 & J013402.53$-$094626.9  & 0.041 & 15.2 & 0.77 & 4287 & 41.1 & 142.6 & 7.46 & 17       & 60 & 40 \\ 	
A03 & J020808.76+003058.8    & 0.084 & 17.7 & 0.74 & 6568 & 40.9 & 110.4 & 7.78 & 77       & 60 & 16 \\ 	
A04 & J075643.71+445124.1    & 0.050 & 15.9 & 0.93 & 3433 & 40.6 & 107.7 & 7.06 & $-$94  & 60 & 19 \\ 	
A05 & J080421.30+100610.9    & 0.034 & 15.6 & 0.59 & 3604 & 40.6 & 134.2 & 7.08 & $-$11  & 60 & 33 \\ 	
A06 & J081755.93+372213.5    & 0.060 & 16.7 & 0.86 & 3693 & 40.5 & 134.0 & 7.09 & $-$157& 60 & 27 \\ 	
A07 & J084137.87+545506.5    & 0.045 & 15.9 & 0.70 & 3202 & 40.9 & 162.1 & 7.14 & $-$26  & 40 & 35 \\ 	
A08 & J230920.26+004523.4    & 0.032 & 14.8 & 0.97 & 5965 & 41.0 & 142.1 & 7.71 & $-$25  & 60 & 36 \\ 	
A09 & J232714.52$-$102317.5  & 0.065 & 15.6 & 0.96 & 2226 & 40.9 & 111.5 & 6.85 & $-$58  & 60 & 23 \\ 	
A10 & J230649.78+005023.4    & 0.061 & 17.0 & 0.86 & 1849 & 40.7 &  93.1 & 6.56 & 125     & 60 & 15 \\ 	
\bottomrule
\end{tabular}
\tabnote{Columns: (1) object ID. (2) SDSS Name. (3) Redshift measured based on stellar lines. 
(4) g-band magnitude from SDSS DR7.  
(5) minor-to-major axis ratio from the KIAS VAG catalog (\url{http://astro.kias.re.kr/vagc/dr7/}) by \citet{Choi+00}.
(6) FWHM of the broad \ha\ emission line measured from SDSS spectrum  by \citet{Eun+17}.
(7) Luminosity of the broad \ha\  emission line from \citet{Eun+17}. 
(8) Stellar velocity dispersion measured from SDSS spectrum.
(9) Black hole mass  derived by Eq.~(\ref{eq_mbh}) using column (6) and (7). 
(10) Slit position angle. 
(11) Exposure time. 
(12) Signal-to-noise ratio at 5100 \AA\ continuum in the central spectrum extracted with a 3\arcsec\ aperture.

}
\end{table*}


\begin{figure*}[htp]
\centering
\includegraphics[width=0.8\textwidth]{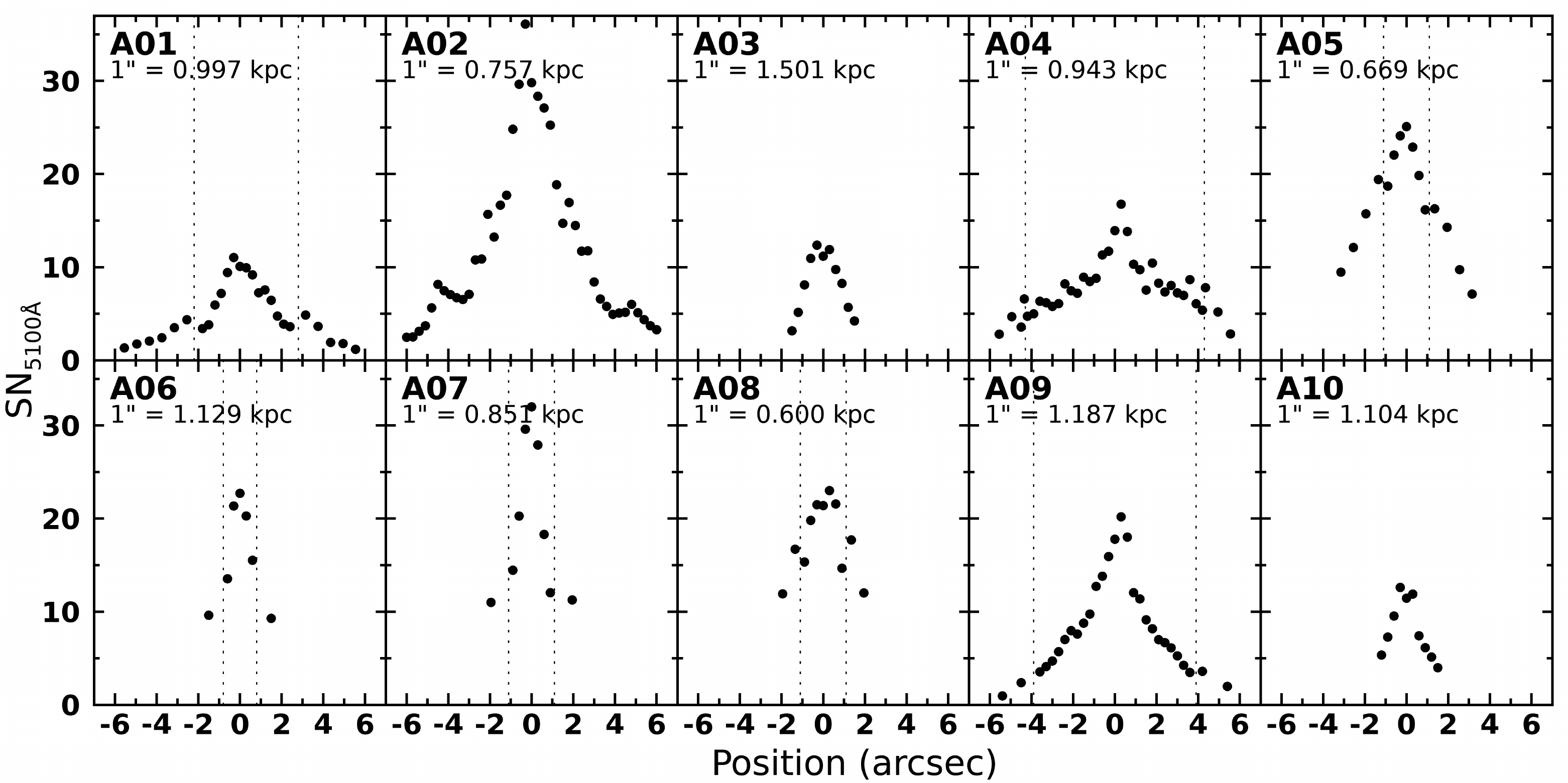}
\caption{Radial distribution of the S/N at 5100\AA\ of each spectrum extracted along the slit direction. 
For the central part within the vertical dotted lines, we used one pixel to extract a spectrum, 
while we combined $3-6$ pixels to extract a spectrum for the outer part
to increase the S/N. The angular scale for each galaxy is labeled in each panel. 
\label{sn}}
\end{figure*}

\subsection{Observations and Data Reduction\label{obs}}

We used the Blue Channel Spectrograph at the 6.5-m MMT (Multiple Mirror Telescope) with a long slit mode.
The slit size was fixed at 1\arcsec$\times$180\arcsec.
The size of the blue channel detector is 2688$\times$512 with pixel size of 15$\times$15 microns.
We chose the 600 line mm$^{-1}$ grating, covering a spectral range 4282$-$5606 \AA, where
the major AGN emission lines, i.e., \hb\ and \oiii\ as well as stellar absorption lines are located.
The spectral setup provided  a spectral resolution of FWHM $\sim $1.45 \AA, corresponding to a line dispersion resolution $\sigma_{inst}\sim37$ \kms\ at the wavelength of \oiii 5007.
The spectral and spatial scales were 0.5 \AA~pixel$^{-1}$ and  0.3 arcsec pixel$^{-1}$, respectively.

We observed ten hidden type 1 AGNs on Nov. 7, 2015. The sky condition during the observing run was good with
typical seeing $\sim$0.7--0.8\arcsec. Each target was observed at airmass $\sim$1.1--1.5 with an exposure of 40 or 60 minutes,
depending on the apparent magnitude. 
In advance we determined the position angle (PA) of the slit along the major axis of host galaxy,
except for one target (A08), for which the slit was positioned along the minor axis to avoid a prominent bar structure. 
Most targets are close to face-on while  one target (A01) is an edge-on galaxy. 
In Figure~\ref{slit} we present the SDSS image of each target along with the slit position.
Table~\ref{target} lists the target properties, slit position angle, and exposure time.
We observed eight G \& K type stars as velocity templates.
A spectrophotometric standard star, BD+28d4211, was also observed for flux calibration.


We processed the data using {\it IRAF} package\footnote{IRAF (Image Reduction and Analysis Facility) is distributed by the National Optical Astronomy Observatories (NOAO).}, by following the standard long-slit data reduction procedure, i.e., bias subtraction, flat fielding,  and wavelength and flux calibrations. Then, we combined multiple exposures to generate a 2-dimensional spectral image for each target. 

From the flux-calibrated 2-dimensional spectral images, we extracted 1-dimensional spectra using a series of one pixel 
(i.e., 0.3\arcsec) aperture size along the slit (spatial) direction, in order to investigate gas and stellar kinematics as a function of radius.
This one-pixel aperture size is corresponding to 200 pc (for A08) $\sim$ 450 pc (for A03) range for our targets.  
Spectra from the central region of each galaxy showed strong enough gas emission lines to study gas kinematics, 
while the spectra in the outer region have relatively low S/N, presenting weak or no emission lines. 
When one-pixel-aperture spectrum  in the outer region showed no emission line feature, we combined $3-6$ pixels, which correspond to $0.9-1.8$\arcsec, to increase S/N. Thus, the spatial resolution at the outer region was reduced (see Figure~\ref{sn}).
We also extracted a single-aperture spectrum using a 3\arcsec\ aperture for each target, 
in order to compare gas and stellar kinematics measurements with those based on SDSS spectra.

\section{Analysis\label{measure}}

\subsection{Stellar pseudo-continuum fitting\label{ppxf}}
 
\begin{figure*}[htp]
\centering
\includegraphics[width=\textwidth]{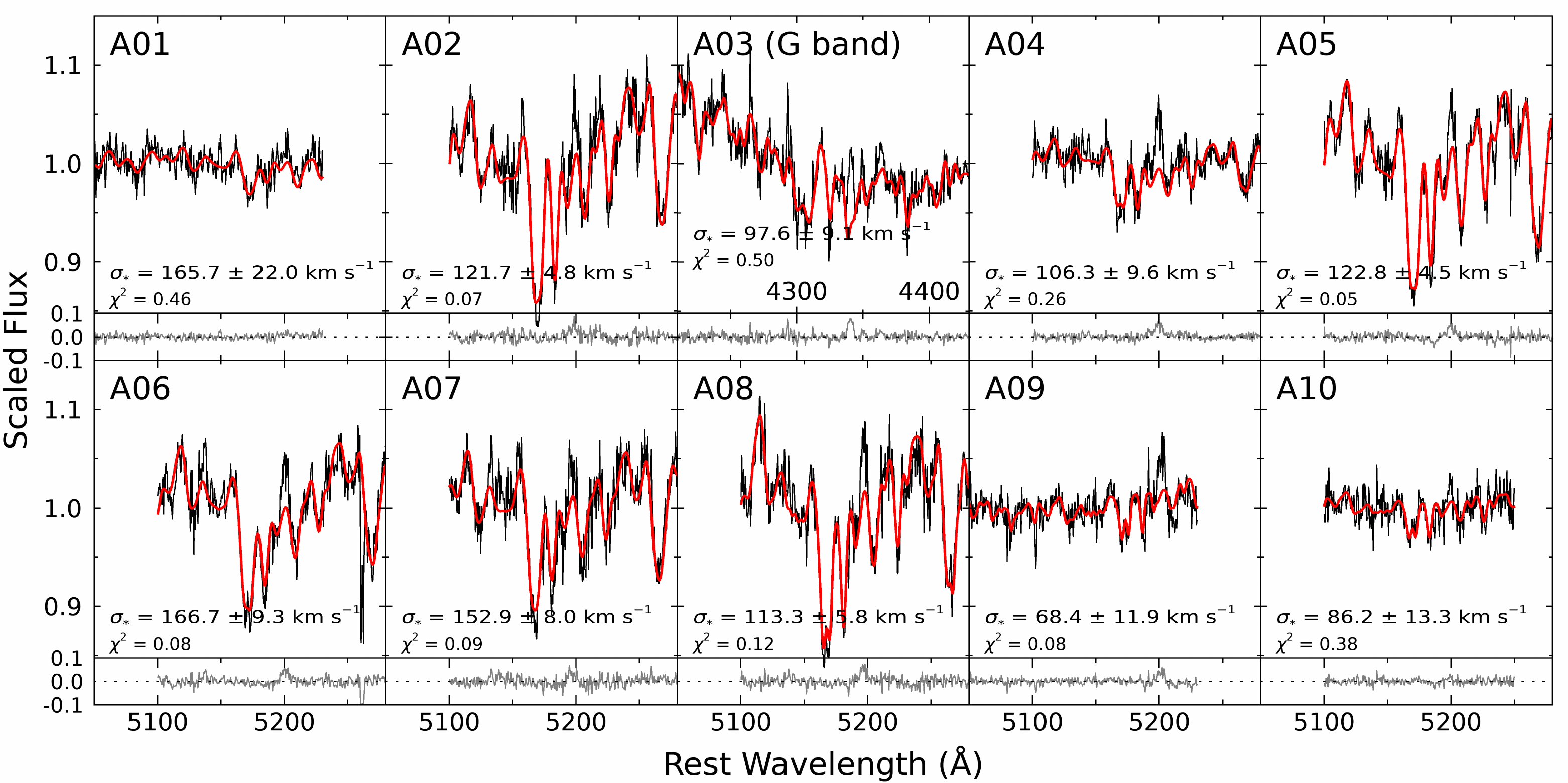}
\caption{Examples of the stellar absorption line fitting. Black line is the spectrum extracted in the central one-pixel aperture size.
Red line is the best-fit continuum model by pPXF. 
Either the Mg b triplet region or the G-band region is used for the fitting process (see Sec. 3.1).
The residual is presented at the bottom of each panel, some of which shows the presence of weak AGN emission lines (e.g. [N I]5200). 
\label{ppxf} }
\end{figure*}

We fitted each spectrum extracted from different positions with stellar models based on the observed K \& G type template stars, 
using the IDL-based the penalized pixel-fitting (pPXF\footnote{http://www-astro.physics.ox.ac.uk/$\sim$mxc/software/}) method \citep{CE04,Cap17}.  
For one galaxy (A03), we used the INDO-US template spectra \citep{Valdes+04} since A03 is the highest redshift target among the sample
and due to the redshift the promising stellar absorption feature is the G-band.
We performed the spectral fitting analysis by fitting continuum in the rest-frame $\sim$4700--5300 \AA\ to generate emission-line spectrum. 
To measure the stellar velocity and velocity dispersion, we utilized a narrower spectral range, e.g.,  
5050--5350 \AA\ around the Mg b triplet or 4210--4430 \AA\ around the G-band.  
These spectral ranges were slightly adjusted for each target according to the redshift of the target.  
We masked emission line regions, especially the \hb\ region, to minimize systematic uncertainty due to the potential broad features generated by AGN emission lines. 
To account for template mismatch, which introduces systematic uncertainty in measuring stellar velocity dispersion, we performed a test using two sets of
templates, i.e., INDOUS and Miles templates, available in the pPXF procedure, for the galaxy spectra extracted from the central 3\arcsec. In this case, various template stars with different spectral types were combined to best match the observed
galaxy spectra. Compared to the stellar velocity dispersion measurements based on our observed K \& G type templates, we found a 0.05 and 0.06 dex (12-15\%)
difference based on INDOUS and Miles templates, respectively. As the systemic uncertainty is relatively small, we use the measurements based on
our observed templates in the following analysis.

For the \ha\ emission line region, where \ha, \nii, \sii\ lines are located, we used the SDSS spectra, which are spatially integrated
within a 3\arcsec\ aperture since our MMT spectra are limited to $\sim$5,600 \AA. 
For the stellar pseudo-continuum fit, we used the MILES templates, which are simple stellar population models with solar metallicity \citep{SB06}. 
After masking out emission lines, we fitted continuum and subtracted the best-fit model around the \ha\ region (6300--6900 \AA) \citep[see][]{Eun+17}.

In Figure~\ref{ppxf}, we show examples of stellar continuum fitting of a central spectrum of each galaxy. 
We adopted the stellar velocity measured from a central spectrum as the systemic velocity of the host galaxy, 
which is then used to calculate the relative velocity shift of gas emission line at each radius.

\begin{figure*}[htp]
\centering
\includegraphics[width=\textwidth]{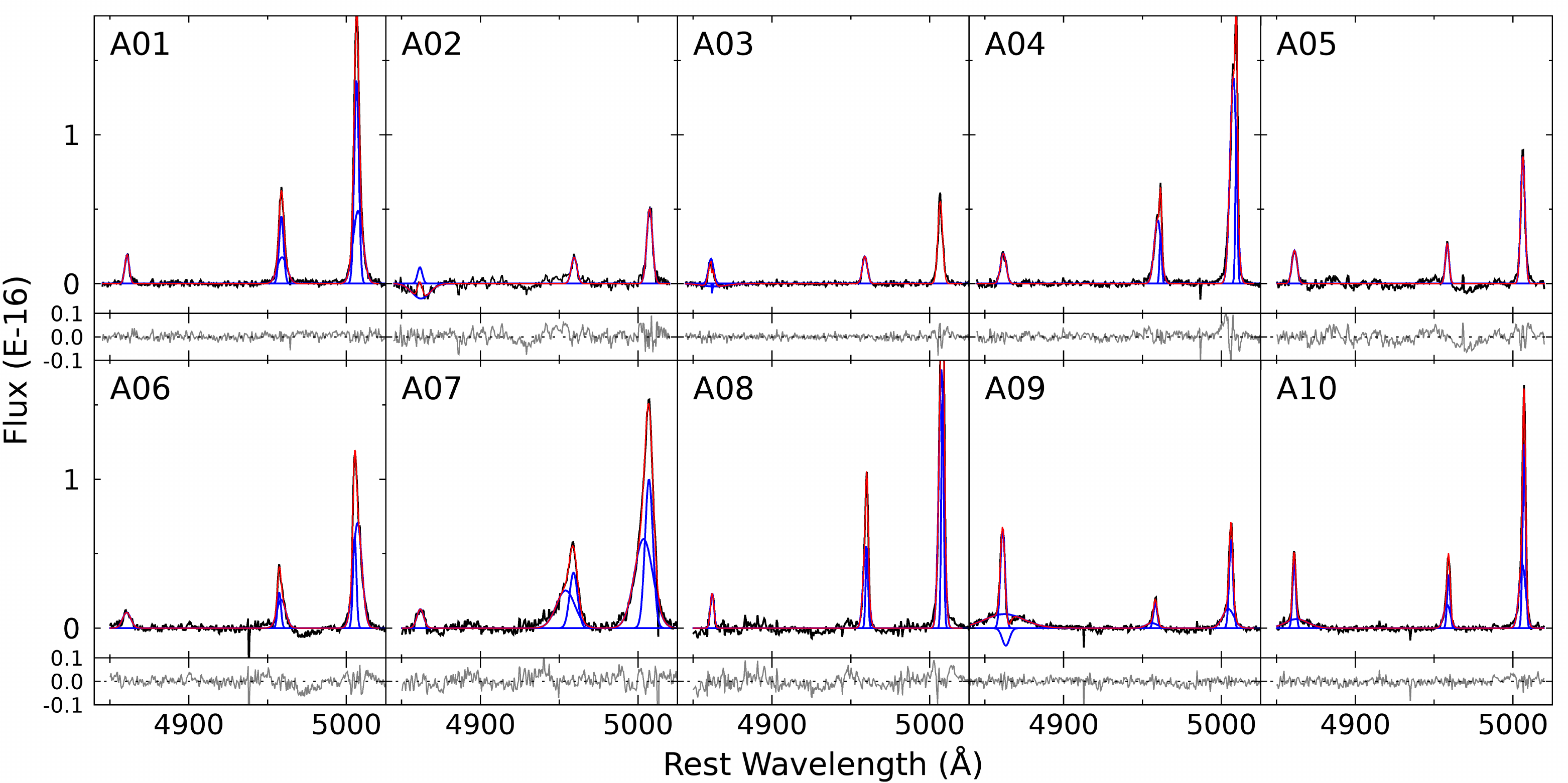}
\caption{Examples of the \hb\ emission line region based on the continuum-subtracted central one-pixel-aperture spectrum (black) and their best-fit models (red). The \oiii\ lines at 4959\AA\ and 5007\AA\ are fitted with either single or double Gaussian model (blue). The residual is presented at the bottom of each galaxy with a grey line. \label{o3fit}
}
\end{figure*}


\begin{figure*}[htp] 
\centering
\includegraphics[width=\textwidth]{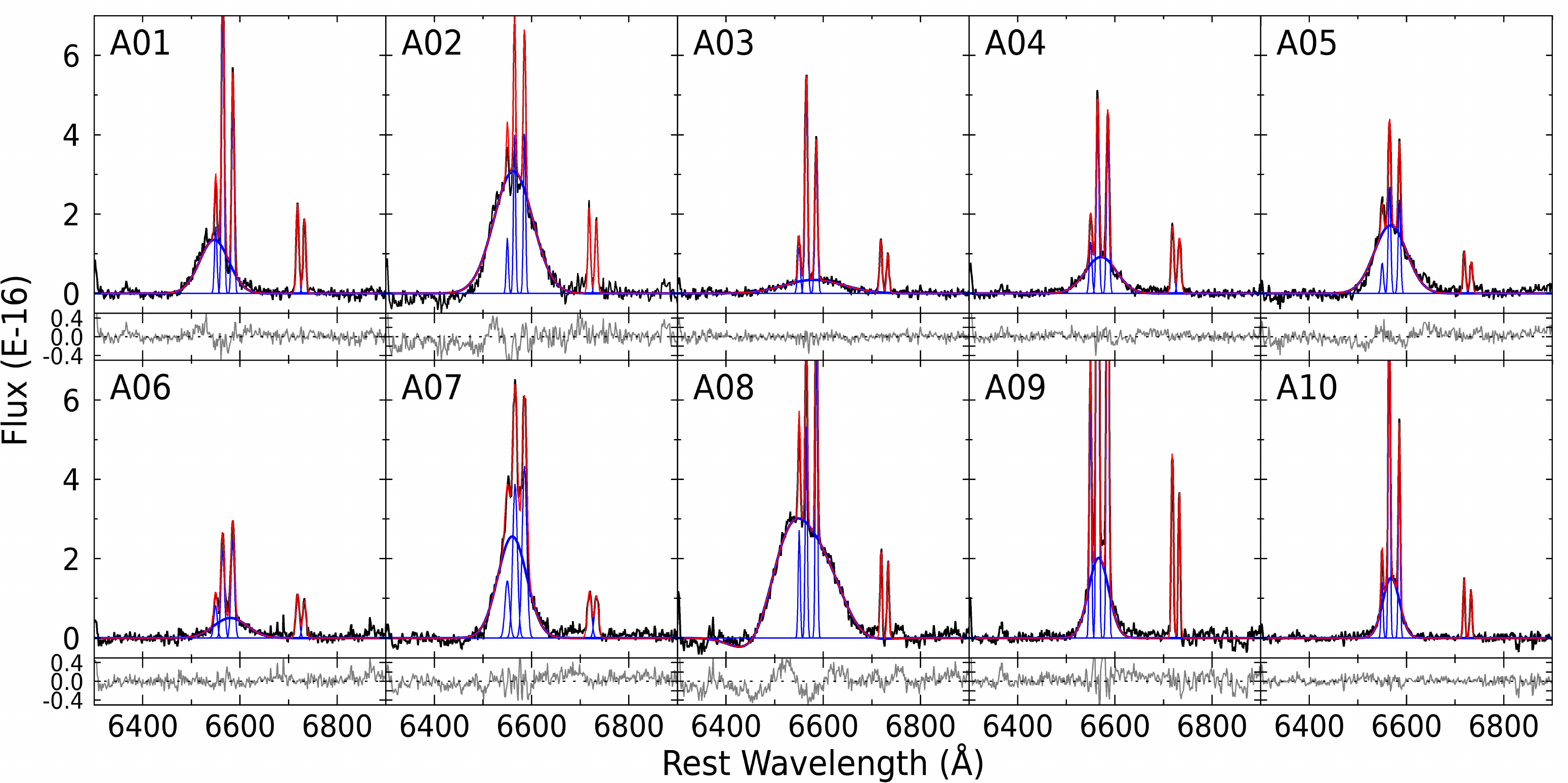}
\caption{\ha\ region of the continuum-subtracted SDSS spectrum (black), overplotted with the best fit model (red), which is composed of single Gaussian models for the narrow emission lines, i.e., \nii, \sii, \ha\ (blue) and a Gaussian model for the very broad \ha\ (FWHM $>$ 1000 \kms). The residual is represented by a grey line. See for details  \cite{Eun+17}.
}
\label{Hafit}
\end{figure*}


\begin{figure*}[htp]
\centering
\includegraphics[width=0.4\textwidth]{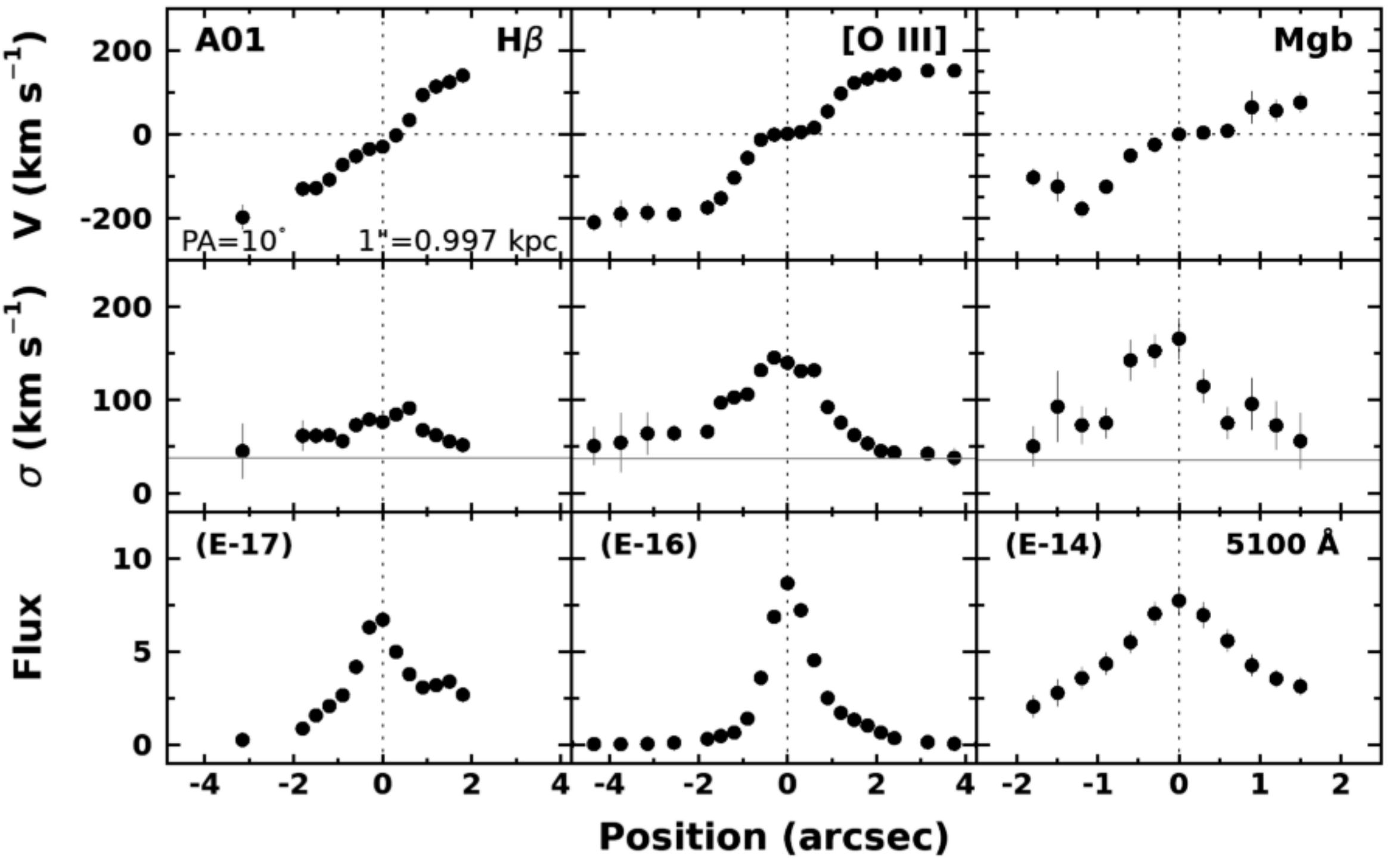}
\includegraphics[width=0.4\textwidth]{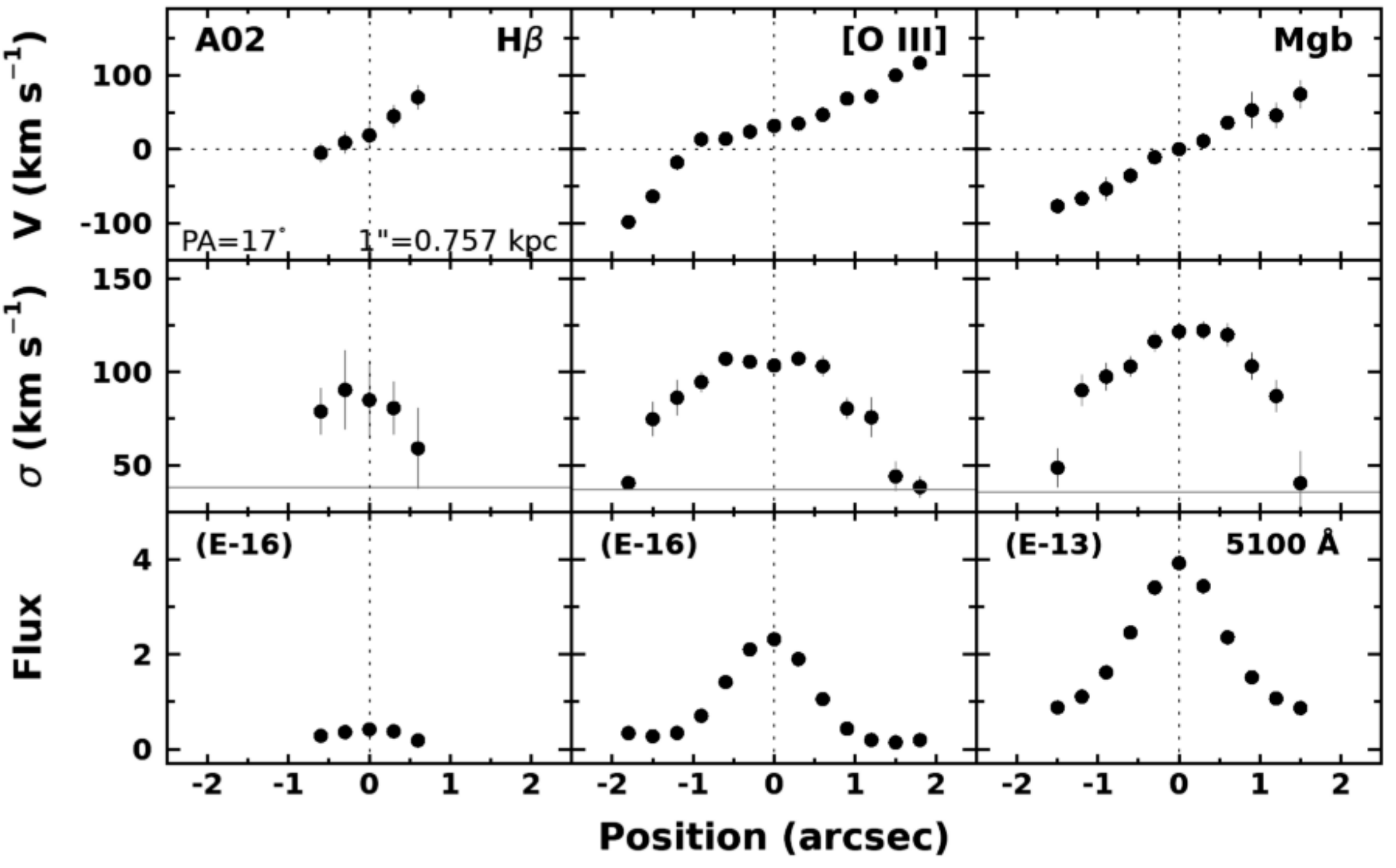}
\includegraphics[width=0.4\textwidth]{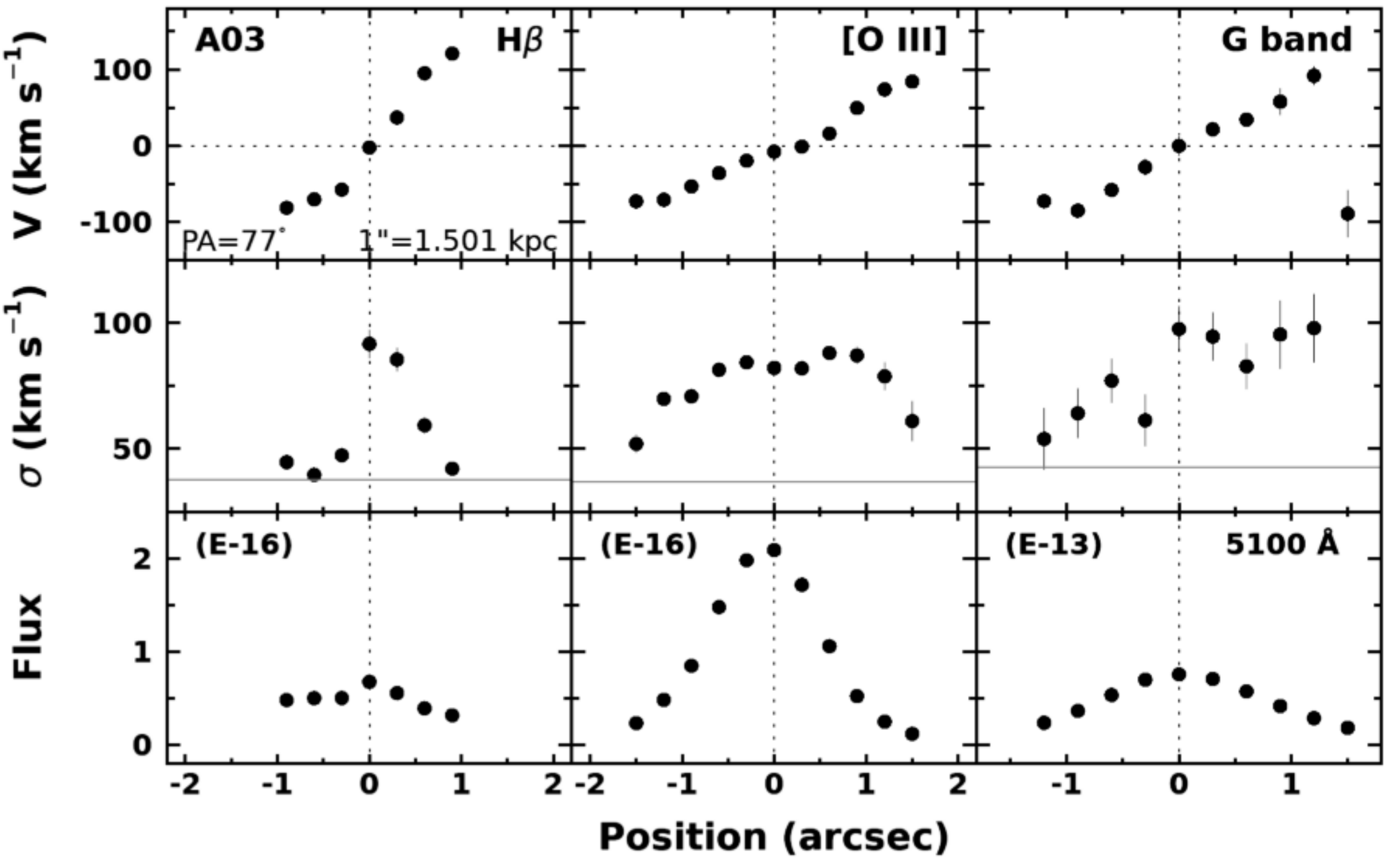}
\includegraphics[width=0.4\textwidth]{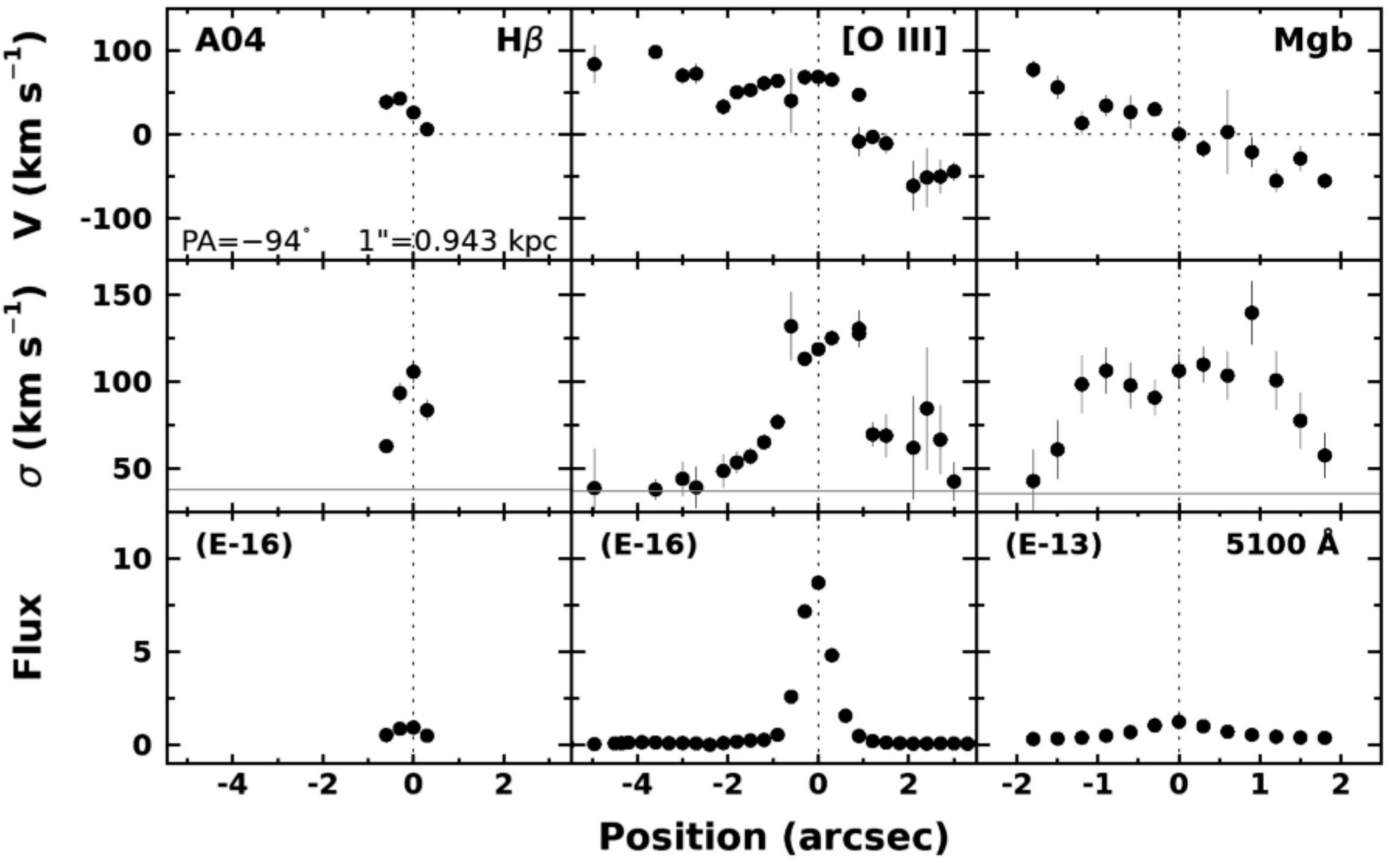}
\includegraphics[width=0.4\textwidth]{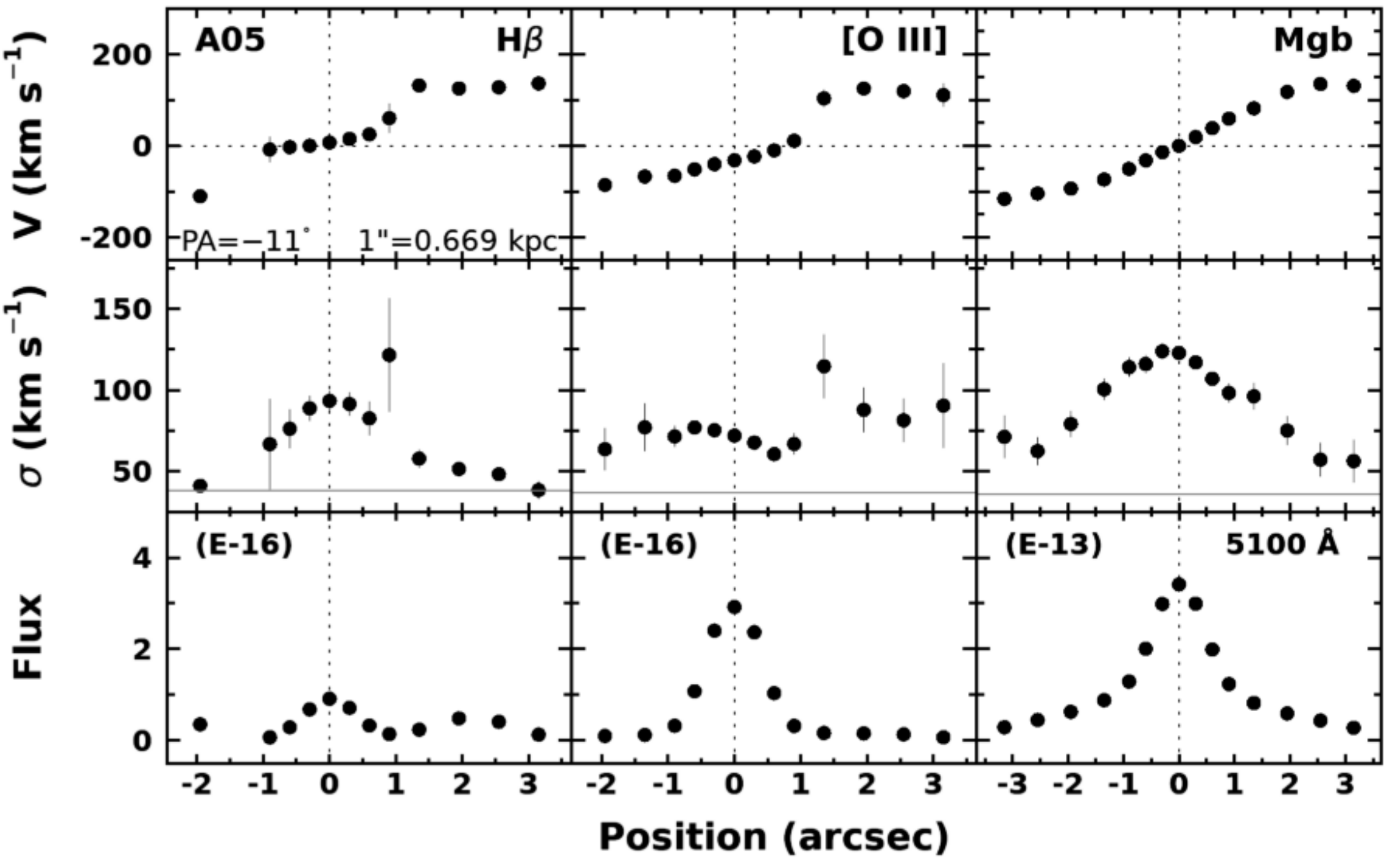}
\includegraphics[width=0.4\textwidth]{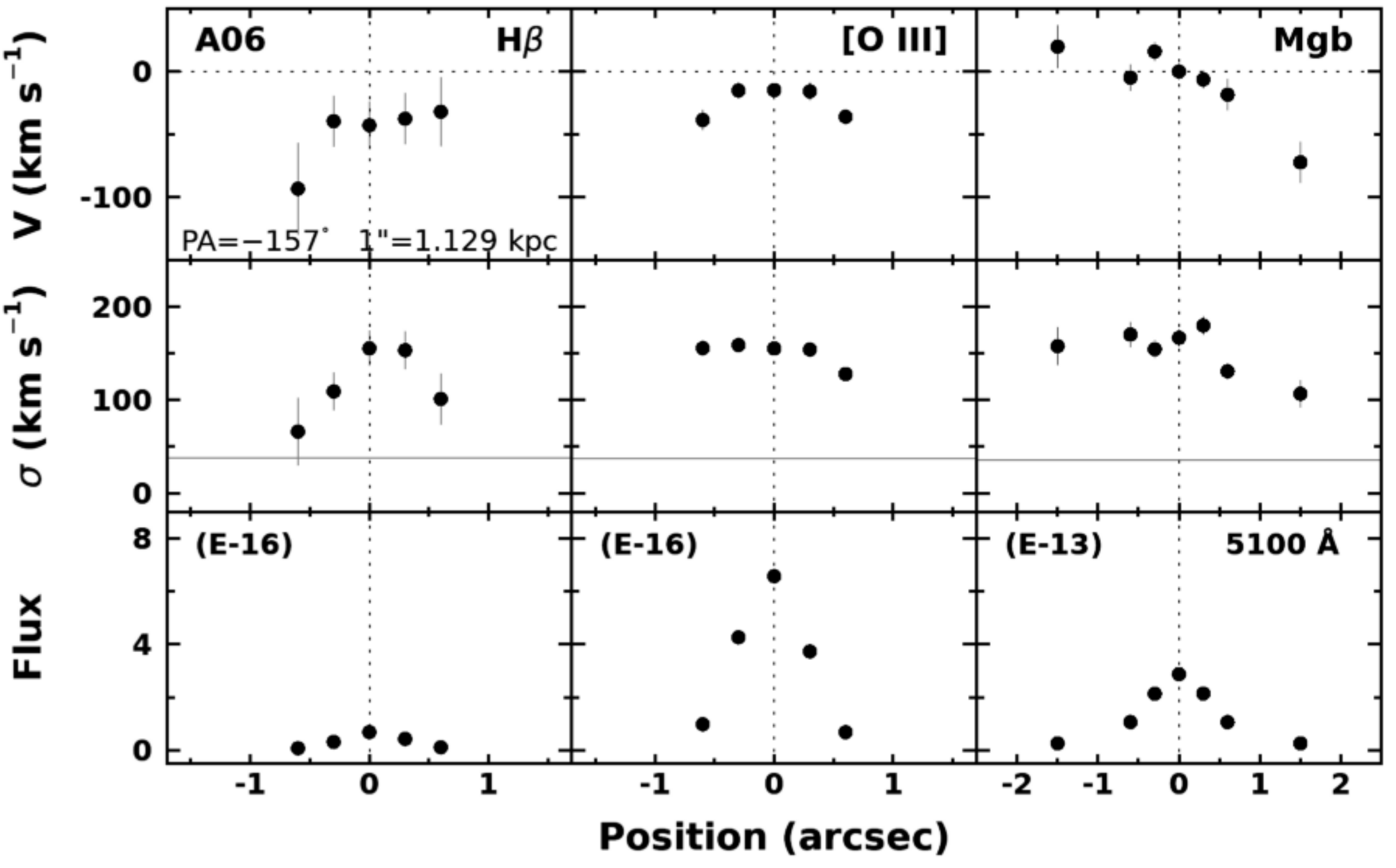}
\includegraphics[width=0.4\textwidth]{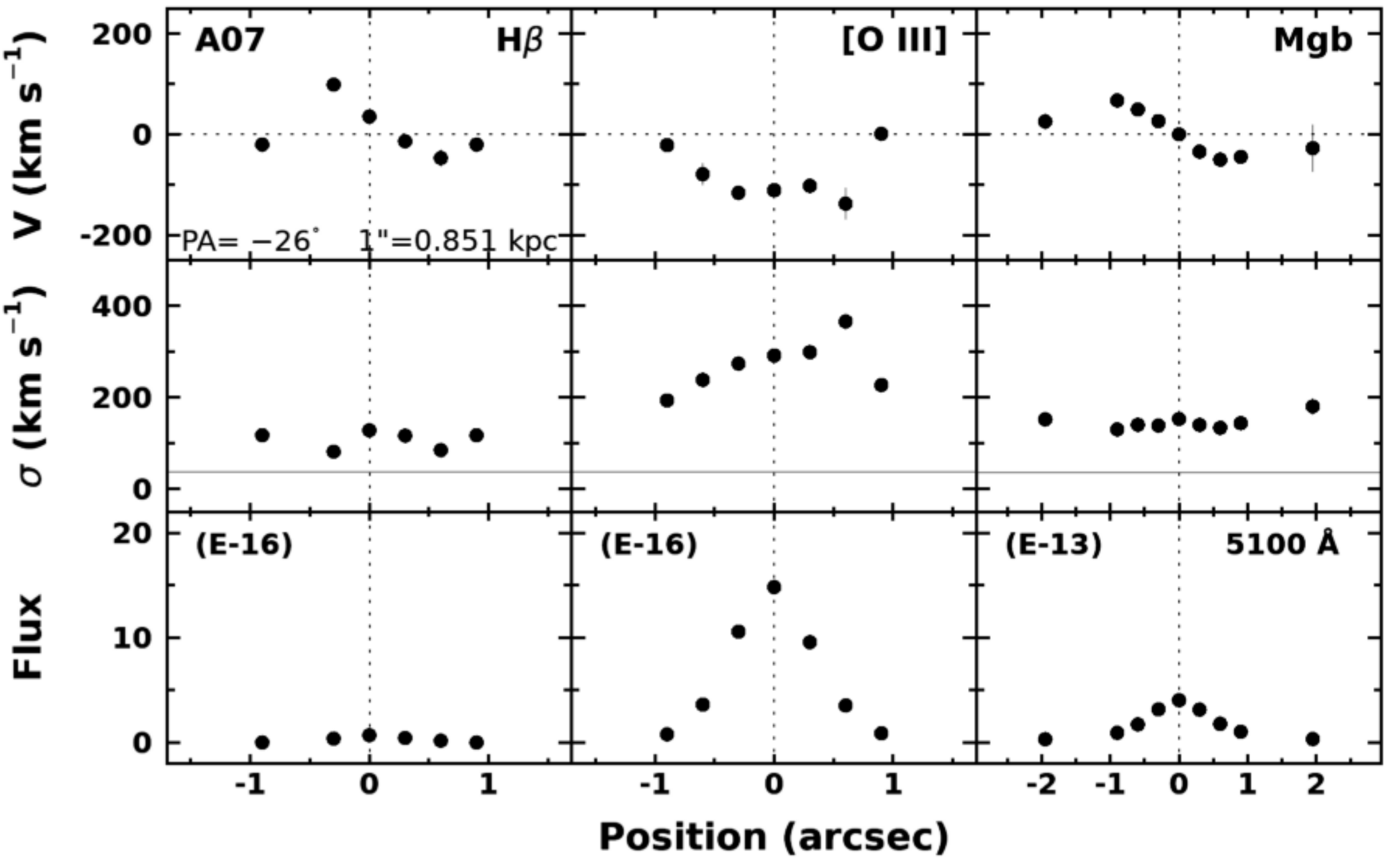}
\includegraphics[width=0.4\textwidth]{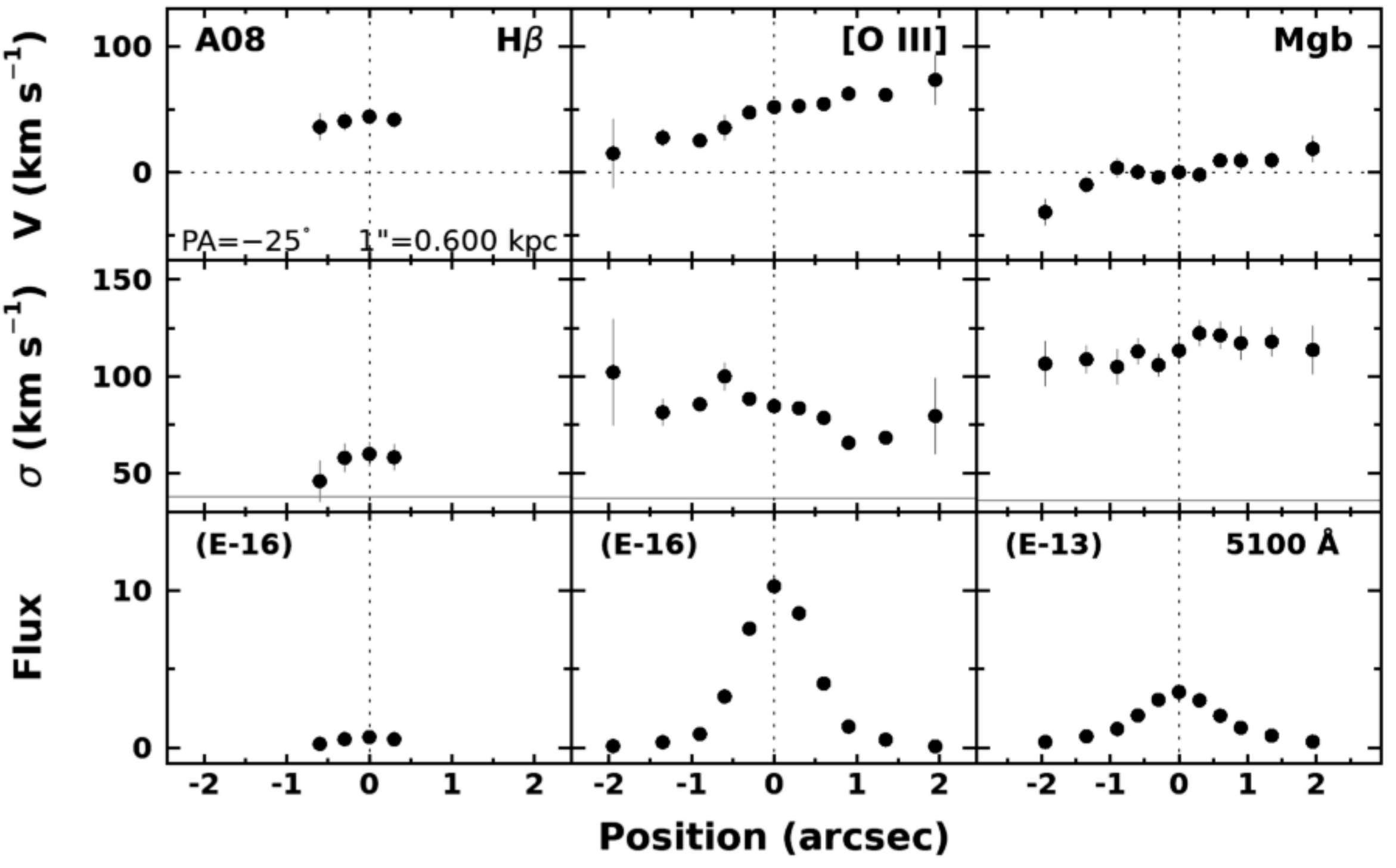}
\includegraphics[width=0.4\textwidth]{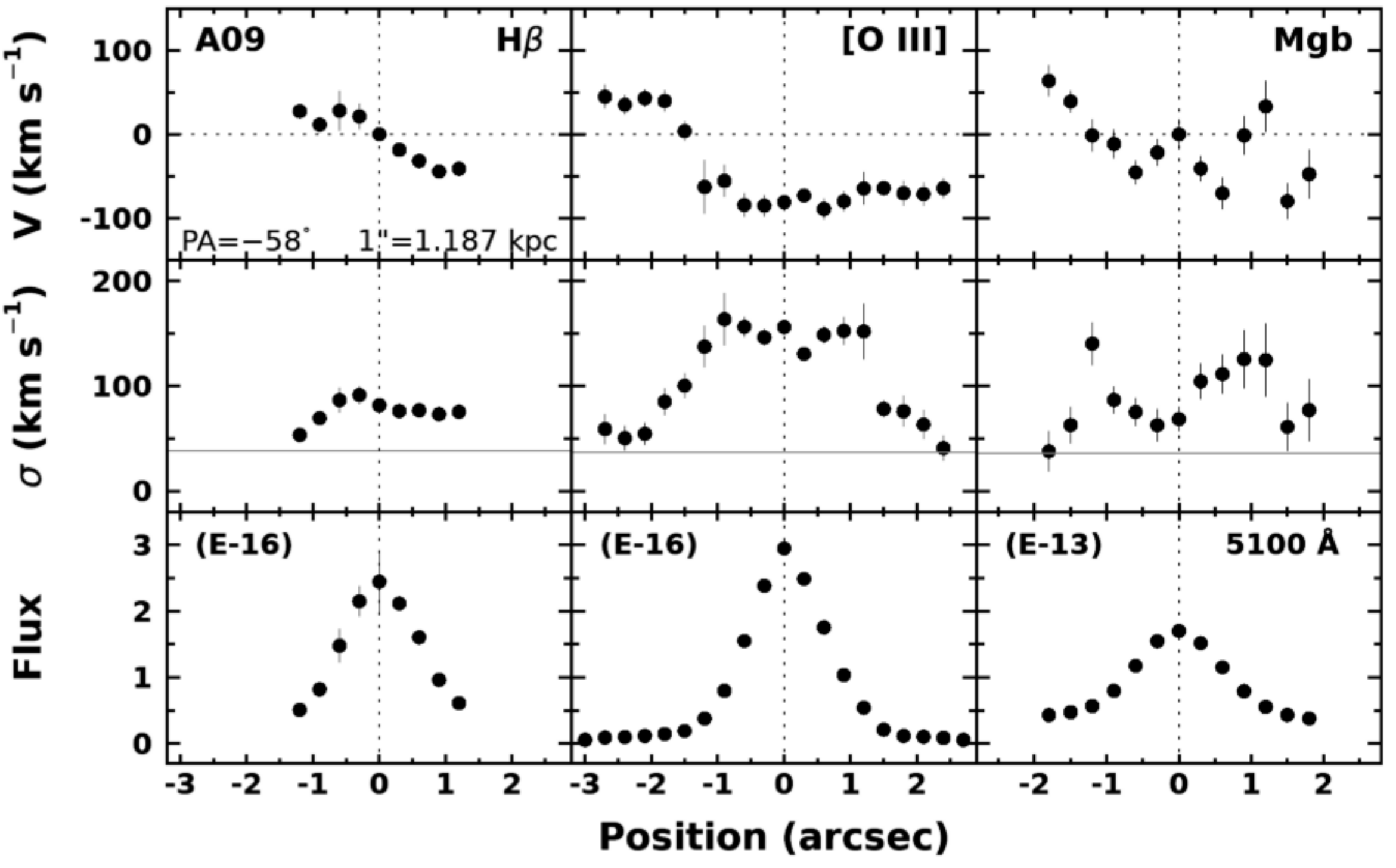}
\includegraphics[width=0.4\textwidth]{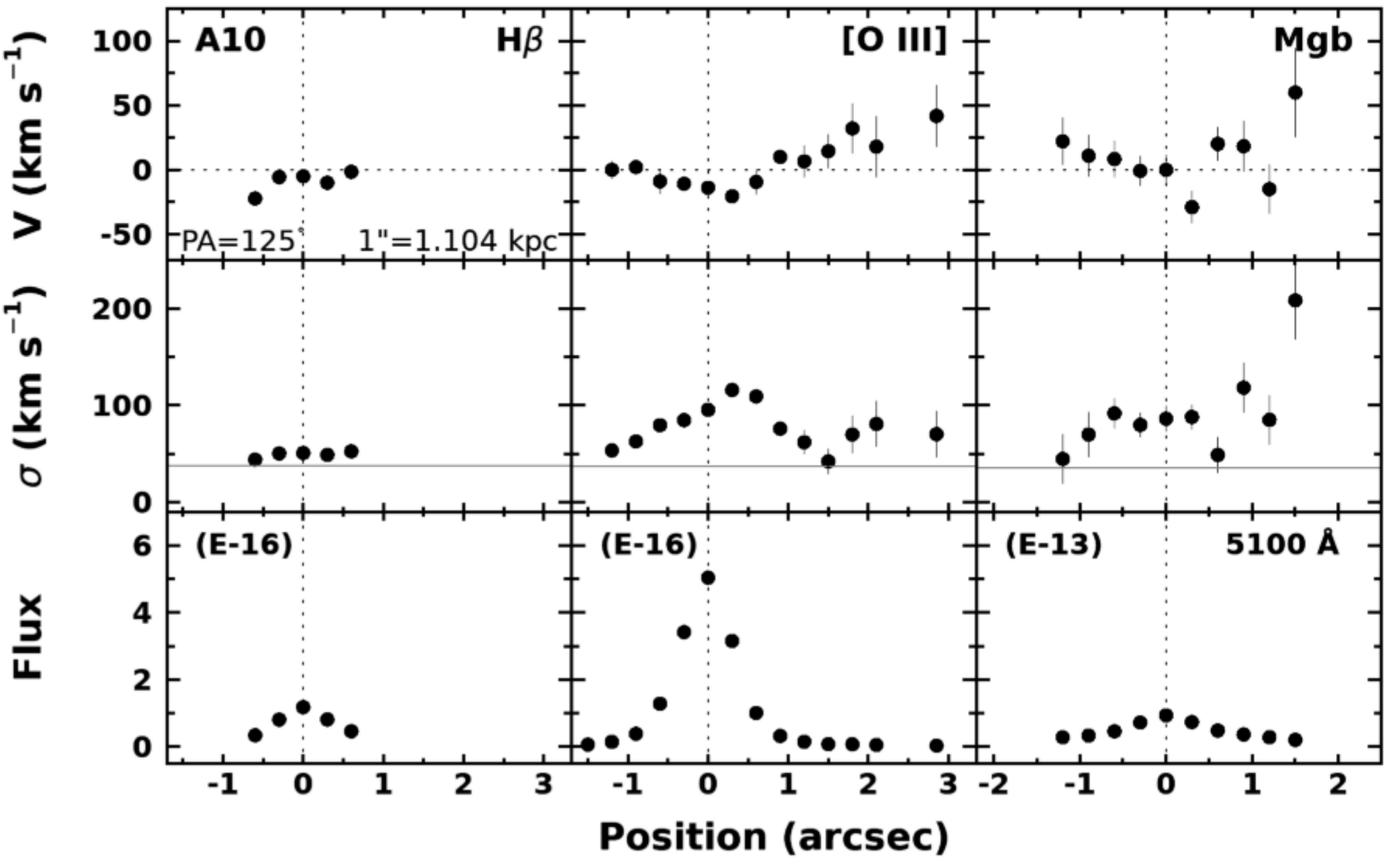}

\caption{Radial distributions of velocity (top), velocity dispersion (middle), and flux (bottom) of \hb\ (left) and \oiii\ gas emission (center), and stars (right) for each target. 
The position angle and the scale are presented in the top-left panel of each object. 
The horizontal lines in the velocity dispersion panels represent the instrumental resolution (36$\sim$43 km s$^{-1}$) at the wavelength of \hb, \oiii, and stellar lines (Mgb or G bands), respectively.  \label{kinematics}}
\end{figure*}

%

\subsection{Emission line fitting}

From the continuum-subtracted spectra, we carried out emission line fitting analysis in the spectral range of 4840--5030 \AA\ for \hb, \oiii$\lambda$4959 and  \oiii$\lambda$5007 simultaneously, in order to measure velocity, velocity dispersion and flux.
We used a single or double Gaussian model using MPFIT\footnote{http://www.physics.wisc.edu/$\sim$craigm/idl/fitting.html} fitting routines  \citep{Ma09} by following our previous analysis \citep[see for more details][]{Woo+16, Woo+17}. 
If \oiii$\lambda$5007 shows a broad wing component, we decomposed the \oiii\ line into a narrow core and a broad wing components.
Note that we only accepted a double Gaussian model, when the peak of the second (wing) component is a factor of three larger than the noise level of the continuum. 
In Figure~\ref{o3fit} we present the examples of the best-fit for \hb\ and \oiii\ emission lines. 
A couple of targets, e.g., A02, A03, and A09 show an absorption feature around \hb, suggesting that the stellar continuum fit was not satisfactory. For these cases, we needed to add a negative Gaussian profile to correct for the absorption feature.
Thus, the gas kinematics measurements based on \hb\ are more uncertain than those based on \oiii.

Based on the best-fit model of each emission line, we calculated the first (i.e., $\lambda_{0}$) and the second moment ($\sigma$, velocity dispersion) of the total line profile as
\begin{equation}
\lambda_{0} = {\int \lambda f_\lambda d\lambda \over \int f_\lambda d\lambda}\\
\end{equation}
\begin{equation}
\sigma^{2}  = {\int \lambda^2 f_\lambda d\lambda \over \int f_\lambda d\lambda} - \lambda_0^{2} ,
\end{equation}
where $f_\lambda$ is the flux density at each wavelength.
From the first moment of \oiii, we calculated the velocity shift with respect to the systemic velocity, which were measured based on stellar absorption lines
in the pseudo-continuum fitting process. We corrected the measured velocity dispersion for the instrumental resolution.

\section{Results\label{result}}

\subsection{Type 1 AGN features}

Based on the \hb\ fitting analysis, we detected a broad component of \hb\ for two targets, A09 and A10,
indicating that these objects are type 1 AGN (Figure~\ref{o3fit}). The FWHM of the broad \hb\ is 1,451 \kms\ and 1,060 \kms, respectively,
for A09 and A10. In the case of the broad \ha, we obtained the FWHM of the broad component as 2,226 \kms\ and 1,849 \kms, respectively for A09 and A10.
Note that the width of \ha\ is  larger than that of \hb. Since the \hb\ with a shorter wavelength range is more susceptible to extinction,
the higher velocity gas at the inner part of the BLR may be more obscured than the case of \ha, truncating the wing component in the observed \hb\ line profile \citep[e.g.][]{GH05}. 
For the other targets, we could not detect the presence of a broad component in \hb, presumably due to 
the overall low flux compared to the noise level of the continuum. 

Six objects, namely, A01, A03, A04, A07, A08, A10 were previously classified as type 1 AGN by  \citet{2006A&A...455..773V}. The classification 
is presumably based on the presence of a broad \ha\ component. For the other four objects, \citet{Eun+17} found a broad component in \ha\ using SDSS spectra
and classified them as type 1 AGN.
In comparison, we present the best-fit results of the \ha\ region, using the continuum-subtracted SDSS spectra in Figure~\ref{Hafit}.
The \ha+\nii+\sii\ spectral features were simultaneously fitted with multi-component Gaussian models by \citet{Eun+17}.  
The clearly detected broad component in the \ha\ line profile indicates that each of these targets is a type 1 AGN. 
The FWHM of the broad \ha\ line ranges from 1,849 \kms\ to 6,568 \kms\ as presented in Table~\ref{target}.
Although we detected a broad component in \hb\ for only two targets, all then AGNs show a strong broad \ha, qualified as type 1 AGNs. 

AGN classification requires a detailed decomposition to determine the presence of a broad component in either \ha\ or \hb\ (or other lines in the near-IR range). When 
AGN-to-stellar flux ratio is relatively low, optical color is much redder than that of quasars or typical Seyfert 1 AGNs, and the broad emission line is weak and
easily hidden by the noisy continuum spectrum. However, these hidden type 1 AGNs are similar to typical type 1 AGNs, albeit with relatively 
low luminosity and low Eddington ratio \citep[see][]{Eun+17}. The black hole mass of our sample ranges from 4 to 60 million \msun\ and the \ha\ luminosity
ranges from 10$^{40.7}$ to 10$^{41.1}$ \ergs. Thus, these AGNs are at the low end of mass and luminosity distribution of local AGNs.

 \subsection{Gas and stellar kinematics}
 
We present the radial distributions of the velocity, velocity dispersion, and flux measured from gas (\hb\ and \oiii) emission lines 
and stellar (Mg b or G band) absorption lines in Figure~\ref{kinematics}.
Note that when the emission line was not resolved due to the limited spectral resolution, we do not show the measurements at that radius in Figure~\ref{kinematics}.


For the majority of the targets, stellar velocity shows a rotation feature, with a typical range of the projected velocity $\pm$100 \kms.
In the case of A09 and A10, the rotation feature is less clear. For A09, this is probably due to the orientation since the target galaxy
is almost face-on while the morphology of the target clearly shows spiral arms. A10 may not have a strong rotation component as expected from the morphology (see Fig.~\ref{slit}). 
The radial profile of stellar velocity dispersion generally shows a peak at the center and a radial decrease, which is a typical 
trend for a gravitational potential of host galaxies while the stellar velocity dispersion at the center is mostly less than 150 \kms.  A few targets (i.e., A04 and A09) show a $\sigma$-drop with a lower stellar velocity dispersion at the very center, which has been reported in the literature for other galaxies \citep[e.g.,][]{Kent+90, Bender+94, Pinkney+03, Kang+13}. 

Gas kinematics represented by \oiii\ and \hb\ are broadly consistent with stellar kinematics, indicating that the motion of ionized gas
is mainly governed by the galaxy gravitational potential. However, some of the targets show somewhat different features due to non-gravitational effects. 
For example, A02, A04, A05, A06, A07, and A08 show a radial velocity profile of gas, which is inconsistent with that of stars, suggesting that that ionized gas
is under the effect of outflows. The radial profile of gas velocity dispersion also shows a different pattern compared to that of stars. Some of these
targets (e.g., A04, A07, A09, and A10) show much higher velocity dispersion of \oiii\ than stars at the center, indicating the presence of non-gravitational effects. 

The different kinematics between gas and stars are expected from the \oiii\ emission line profile since all but the three targets (A02,  A03,  A05) show a broad wing component at the central region in Figure~\ref{o3fit}. The broader wing component of \oiii\ has larger velocity dispersion ($>$ 150 \kms) than stellar lines or the narrow component of \oiii, indicating that outflows change the gas kinematics as similarly detected in various AGN based on spatially resolved and spatially integrated data \\citep[see][]{Woo+16,Woo+17, Le+17, Kang+19, Luo+19, Wylezalek+20}.

In general, gas kinematics can be different from stellar kinematics due to various effects, including outflows, inflows, and galaxy-galaxy interaction.  
Also, star formation can generate significant outflows while the velocity dispersion of ionized gas is modest compared with AGN-driven outflows \citep{Woo+16, Rakshit+18}. 
With the current kinematical features, we do not have strong evidence of outflows. However, gas kinematics is different from stellar kinematics at the very center of host galaxies, where AGN driven outflows are expected, we suggest that the kinematic difference is likely due to outflows.

\begin{figure}[h] 
\centering
\includegraphics[width=0.4\textwidth]{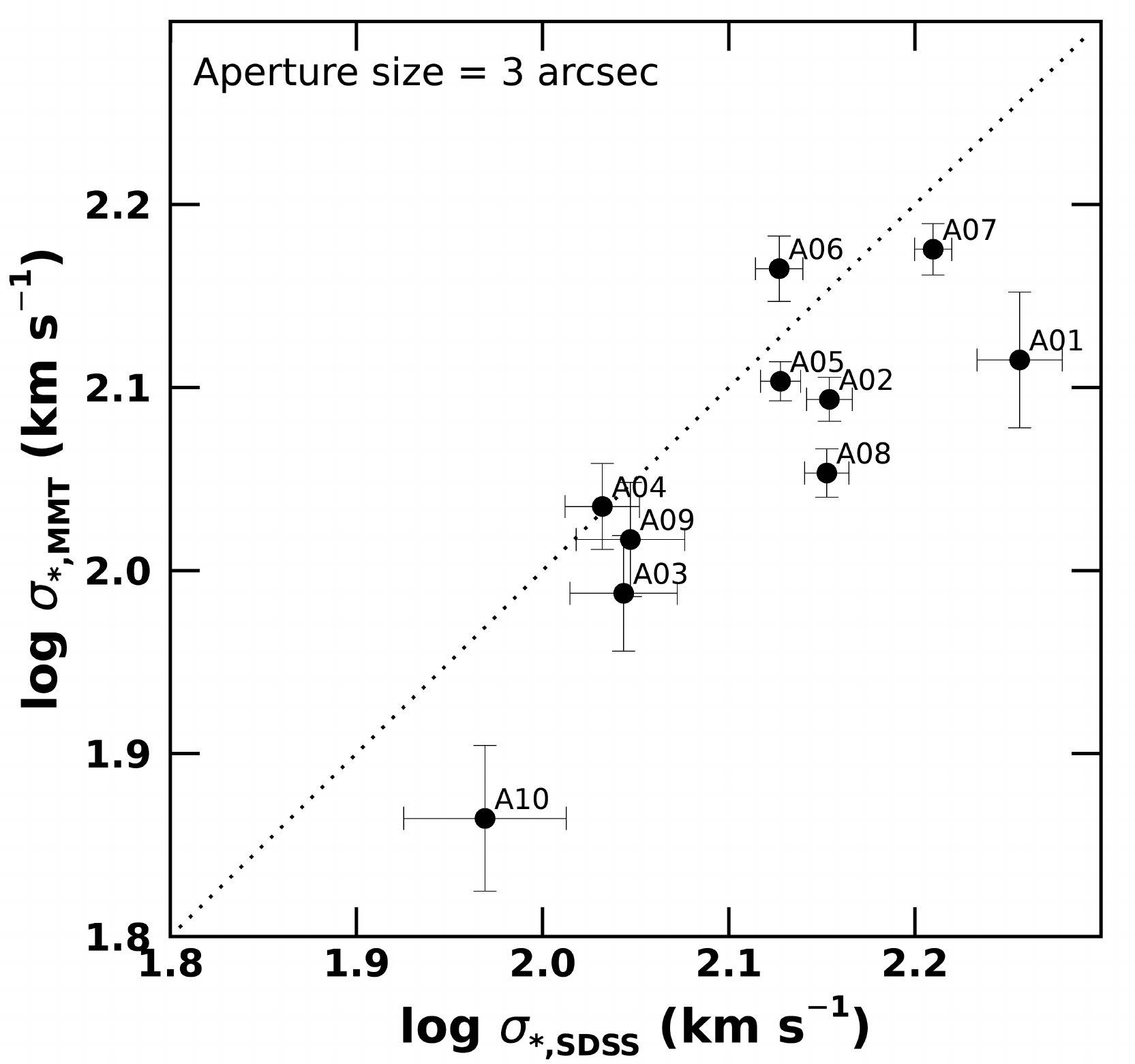}\\
\includegraphics[width=0.4\textwidth]{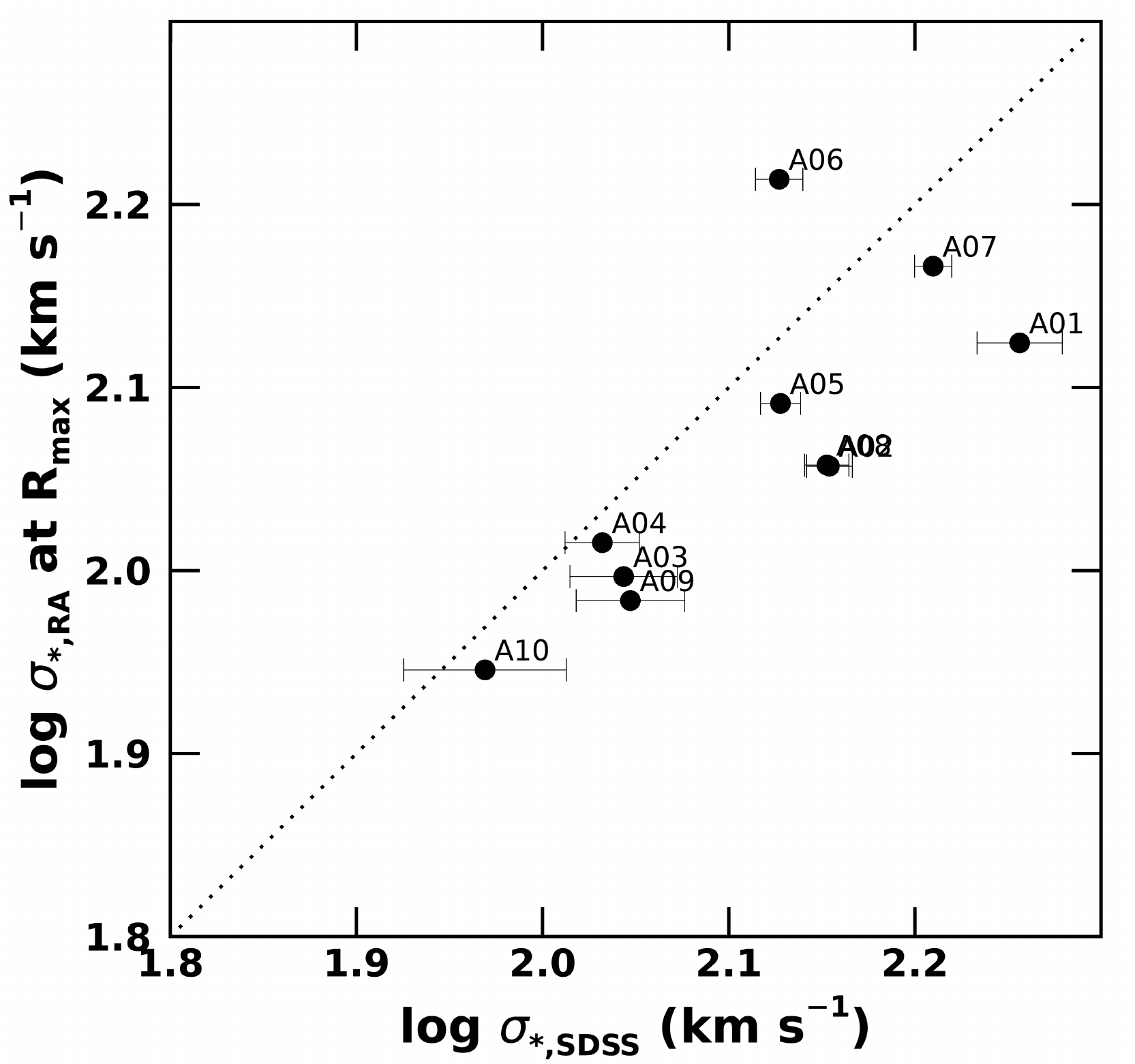}
\caption{Comparisons of stellar velocity dispersions. Top: single-aperture based SVD of the SDSS spectrum and of the MMT spectrum with a 3\arcsec\ aperture. 
Bottom: Comparing the SVD from SDSS with the flux-weighted SVD based on the spatially resolved MMT data.
The dotted line is the 1:1 line for comparison.}  
\label{fig:comp_svd}
\end{figure}

\begin{figure*}[htp] 
\centering
\includegraphics[width=\textwidth]{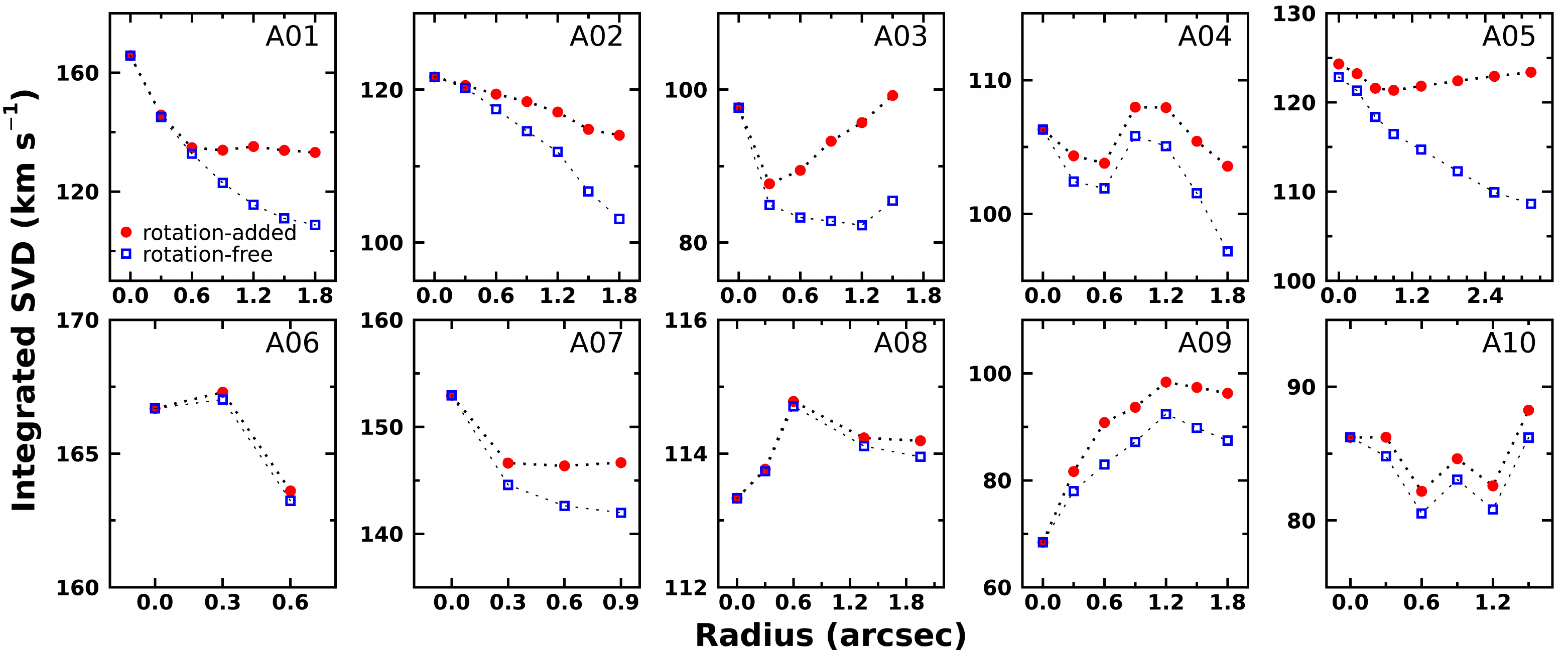}
\caption{Flux-weighted SVD as a function of the integration radius. The SVDs are calculated by including rotation effect (rotation-added based on Eq.~(\ref{eq_RA}); red filled circle) or without adding rotation velocity (rotation-free based on Eq.~(\ref{eq_RF}); blue open square).   
\label{fig:svd}}
\end{figure*}

\subsection{Effect of rotational broadening in SVD\label{svd} }

For late-type galaxies, which are generally rotation-supported systems, stellar velocity dispersion can be easily overestimated if a single-aperture spectrum is used due to the rotational broadening of stellar lines. 
Thus, spatially resolved stellar or gas kinematics are required to properly represent the gravitational potential. 
Based on our spatially resolved long-slit data, we performed several tests to investigate the systematic uncertainty of SVD based on single aperture spectra.

First, we compare stellar velocity dispersion measured from the 3\arcsec-fiber SDSS spectra with that from our MMT spectra, 
which were also extracted with a central 3\arcsec\ aperture in Figure~\ref{fig:comp_svd} (top panel). 
Note that we used 1\arcsec\ slit for the MMT spectra, and the seeing condition was better than that of the SDSS spectra. Thus, the MMT spectra proves more central region than SDSS spectra. On average, the SVD measured from SDSS spectra is 0.05 dex (a factor of 1.12) larger than that of MMT spectra, presumably due to the different amount of the rotational broadening in the extracted spectra. 
The larger aperture of SDSS covers larger part of the rotating disk and therefore it may inflate the SVD measurement.

We calculate the effective stellar velocity dispersion using Eq.~(\ref{eq_RA}), which has been generally used for representing the gravitational potential of target galaxies in comparing with black hole mass \citep{Gul+09, KH13, Bell+14}. 
We also calculated rotation-free ($RF$) stellar velocity dispersion to separate a pressure-supported component from a rotation-supported component, following \citep{Woo+13, Kang+13, Bennert+15}:
\begin{equation}
\sigma^{2}_{*,RF} = { \int^{R_e}_{-R_e}\sigma_{*}(r)^{2}I(r)dr \over \int^{R_e}_{-R_e}I(r)dr }.
 \label{eq_RF}
\end{equation}

To calculate the effective SVD, an effective radius of the bulge component has to be determined as required by Eq.~(\ref{eq_RA}) and (4).
We measured the effective radius of each target, using the SDSS g-band images by using the GALFIT\footnote{https://users.obs.carnegiescience.edu/peng/work/galfit/galfit.html} software \citep{Peng+02, Peng+10}.
In this fitting, galaxy image was modeled with a combination of a PSF model, one S\'ersic profile \citep{Ser63} with $n=4$ (classical bulge), and additional $n$ as a free parameter if necessary. The central region of seven targets were modeled with a $n=4$ component, while three targets (A04,A05, A09) were modeled with {\it n}$\sim$0.5, $\sim$1.4, and $\sim$1.9, respectively, suggesting that the bulge is not a classical bulge. 
However, the effective radii from GALFIT are larger than the region where we spectroscopically determined stellar kinematics based on the MMT data. 
Thus, we cannot determine flux-weight effective SVDs within the effective radius. Alternatively, we used the maximum radius, within which we obtained stellar velocity and velocity dispersion (Figure~\ref{fig:svd} and Table~\ref{Reff}). Note that due to this reason, the presented effective SVD may not properly represent the galaxy potential.
However, with the limited spatial coverage, we can test the effect of rotation at the central part of the host galaxies.

In Figure~\ref{fig:comp_svd} (bottom panel), we compare the effective SVD based on the spatially resolved data from MMT with the SVD based on the single-aperture SDSS spectra. We find that SDSS-based SVD is on average 0.05 dex larger than that from spatially resolved data as expected. 
The effect of the rotational broadening depends on the inclination of the host galaxy to the line-of-sight. As the target galaxies are mostly face on, the rotational broadening is relatively weak. However, for more inclined galaxies, the effect of the rotational broadening in a single aperture spectrum is expected to be larger. In fact, we find that the edge-on galaxy, A01 showed the largest discrepancy between the spatially resolved SVD and the single-aperture based SVD. 

To further investigate the effect of the rotation in determining effective SVDs, we present the effective SVD as a function of the integration size with and without adding rotation component based on Eq.~(\ref{eq_RA}) and Eq.~(\ref{eq_RF}), respectively (Figure~\ref{fig:svd}). 
Because the radial profile of SVD generally peaks at the center, the effective SVD becomes smaller as we use a larger integration radius. 
If we use Eq.~(\ref{eq_RF}) to calculate rotation-free SVD, this trend is strong and the spatially-integrated SVD directly represent the radial profile of SVD presented in Figure 6. 
In contrast, if we calculate rotation-added SVD, the decrease of SVD with increasing
integration size becomes much weaker since the effect of the rotation component becomes stronger at outer regions, thus, compensating the decrease of velocity dispersion. This effect is most strikingly detected in A01 (see also A05), which is an edge-on disk galaxy with high rotation velocity while SVD is much weaker (below 100 \kms\ even at the center). 
At the largest integration size of A01, the rotation-added SVD is 133.2 \kms\ while the rotation-free SVD is 108.8 \kms. 
In contrast, A02 does not show such a strong difference since the galaxy is close to face-on and the projected rotation velocity is relatively low (see Figure~\ref{kinematics}). Note that effective SVD calculated based on Eq.~(\ref{eq_RA}) can significantly depend on the inclination of target galaxies. In other words, rotation-added SVD can be different for twin galaxies
if their inclination angle to the line-of-sight is different.


\begin{table}[h]
\centering
\caption{Spatially resolved measurements. \label{Reff}}
\begin{tabular}{rrrrr}
\toprule
ID & r$_{e}$ & r$_{max}$ &  $\sigma_{\rm *,RA}$ & $\sigma_{\rm *,RF}$    \\
      & (\arcsec)  & (\arcsec) & (\kms) & (\kms)  \\
(1) & (2) & (3) & (4) & (5)     \\
\midrule
A01 & 3.9 		& 1.8 & 133.2 & 108.8  \\	
A02 & 7.4 		& 1.8 & 114.0 & 103.1  \\ 	
A03 & 1.8 		& 1.5 &   99.2 &  85.5  \\ 	
A04 & 7.5 		& 1.8 & 103.6 &  97.2  \\ 	
A05 & 4.8  	& 3.2 & 123.4 & 108.6  \\ 	
A06 & 4.8 		& 0.6 & 163.6 & 163.2  \\ 	
A07 & 10.3 	& 0.9 & 146.7 & 142.0  \\ 	
A08 & 2.3		& 2.0 & 114.2 & 114.0  \\ 	
A09 & 6.1   	& 1.8 &   96.3 &   87.4  \\ 	
A10 & 7.0   	& 1.8 &   88.2 &   86.2  \\ 	
\bottomrule
\end{tabular}
\tabnote{Columns: (1) object ID. 
(2) Effective radius from GALFIT. 
(3) Available maximum radius in Fig.~\ref{fig:svd}.
(4) Rotation-added SVD with a maximum integration radius.
(5) Rotation-free SVD with a maximum integration radius.  
\label{Reff}}
\end{table}

\subsection{\msigma\ relation\label{msigma} }

We investigate the black hole mass (\mbh) correlation with host galaxy SVD (\msigma) relation, 
using the determined SVDs based on spatially resolved data from our observation as well as single-aperture 
spectra from SDSS. First, we determine black hole mass for each object, using the broad \ha\ emission line. 
AGN black hole masses can be determined based on the virial assumption of the BLR gas. 
Since we do not have a long-term monitoring data, 
we adopt the BLR size - luminosity relation based on the sample of the reverberation-mapped AGN \citep{Bentz+13}. 
For the sample of hidden-type 1 AGN, we used the single-epoch
mass estimator, which is calibrated for the luminosity and FWHM of the \ha\ emission line by \citet{Woo+15}:
\begin{eqnarray}
\nonumber
\rm {M_{BH}} ={\it f}  \times 10^{6.544} \left ({L_{\rm H\alpha} \over 10^{42}~ \rm{ erg~s}^{-1} }\right )^{0.46}  \\
 \times \left({\rm{FWHM}_{H\alpha} \over 10^{3}~ {\rm km~s}^{-1}}\right )^{2.06} M_\odot ,
 \label{eq_mbh}
\end{eqnarray}
where $f$ is a virial factor and we used the empirically determined log $f = 0.05$ for the FWHM-based \mbh\ estimator \citep{Woo+15}.   
L$_{\rm H\alpha}$ and  FWHM$_{\rm H\alpha}$ are the luminosity and the width of the broad \ha\ emission line, respectively. 
Note that we utilized the SDSS spectra for measuring the flux and FWHM of the \ha\ emission line since
the \ha\ line is not spatially resolved and the flux calibration of the SDSS spectra is more reliable. 
We list the derived \mbh\ of each target by \citet{Eun+17} in Table~\ref{target}.

Then, we investigate whether the hidden type 1 AGN follow the \msigma relations, which are defined by more massive
black holes in inactive and active galaxies. Although there are various versions of the \msigma\ relation, depending on
the sample and the method of black hole mass determination, we use the best-calibrated \msigma\ relation based on
the joint-fit using the combined sample of inactive galaxies with dynamical black hole masses, and active galaxies
with reverberation-based black hole masses by \citet{Woo+15}. 

\begin{figure}[t]
\centering
\includegraphics[width=0.45\textwidth]{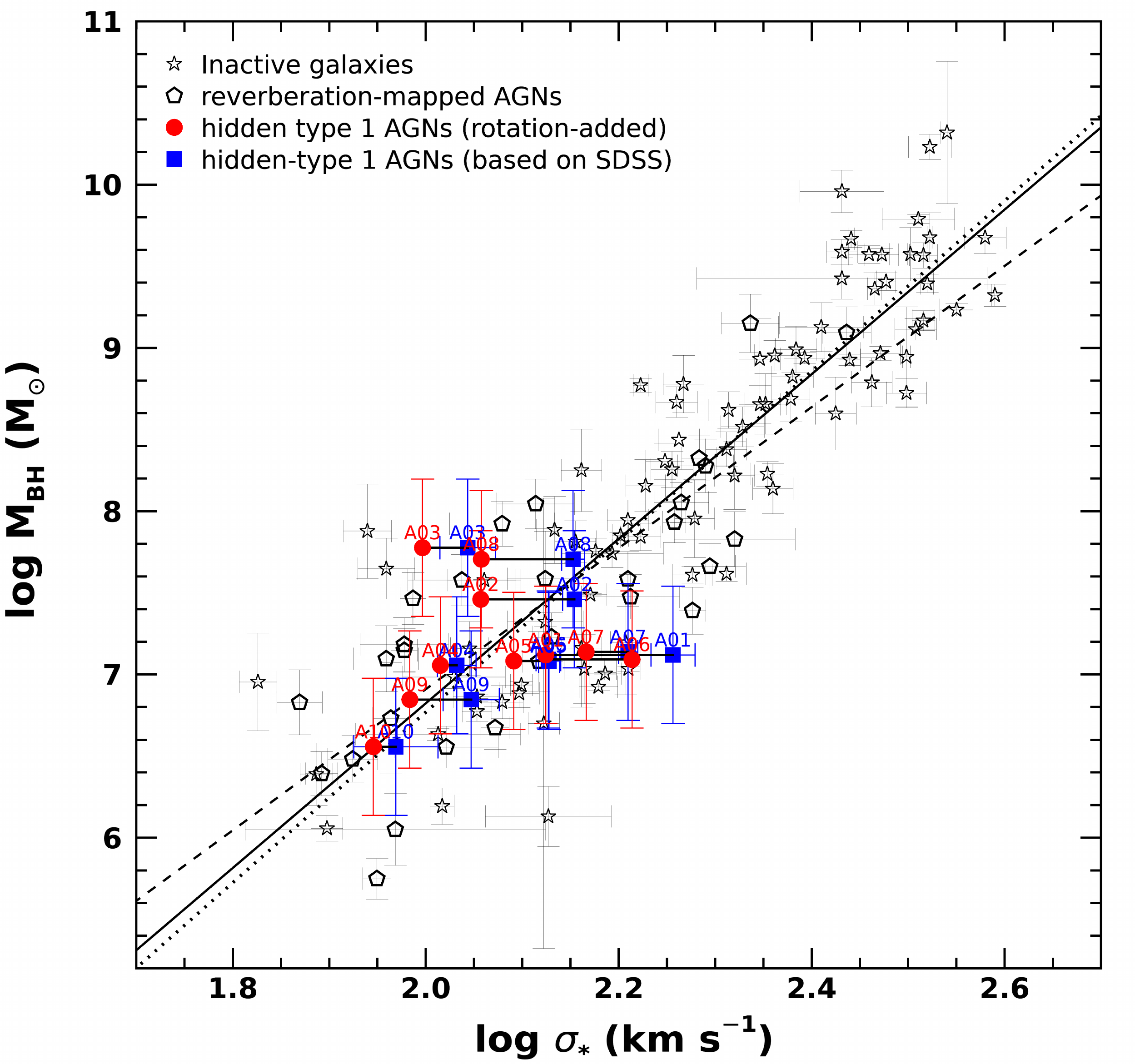}
\caption{\msigma relation of hidden type 1 AGNs compared to inactive galaxies (open stars) and reverberation-mapped
AGN (open pentagon). The best fit \msigma\ relation is represented by a solid line based on the join-fit for inactive galaxies
and reverberation-mapped AGN sample, a dotted line based on 84 inactive galaxies, 
and a dashed line based on the 29 reverberation-mapped AGNs \citep[see for details][]{Woo+15}. 
For the hidden type 1 AGNs, we present both the  effective SVD (filled red circle) 
and the SVD based on single aperture SDSS spectra (blue filled square) for a given target, which are connected with a black line. 
 }
 \label{fig:msigma}
\end{figure}

Using the effective SVDs based on our spatially-resolved kinematics measurements, we investigate the \msigma\ relation by comparing hidden type 1 AGN with inactive galaxies and reverberation-mapped AGNs in Figure~\ref{fig:msigma}. Note that SVDs of reverberation-mapped AGNs are based on single-aperture spectra while those of inactive galaxies are obtained from spatially-resolved kinematics studies.
First, we compare the \msigma\ relation using the rotation-added SVD. 
Our sample of hidden type 1 AGN show an average offset of  $0.03 \pm 0.53$ dex  (in black hole mass) from the best-fit relation (solid line), 
indicating that hidden type 1 AGNs follow the relation, albeit with a significant scatter. 
As a comparison, we used the rotation-free SVD defined with Eq.~(\ref{eq_RF}), in order to investigate the effect of the rotational component in the host galaxy. 
In this case, we find a small offset with a large uncertainty as $0.21 \pm 0.58$ dex,  which may imply that black hole is more massive than expected from the \msigma\ relation. 

Instead, if we use the single-aperture based SVD from SDSS, the offset becomes $-0.20 \pm 0.48$ dex. As expected from the direct comparison 
of the spatially resolved and single-aperture SVDs in Figure 7, the hidden type 1 AGNs show a negative offset (i.e., smaller black hole mass or larger SVD)
from the \msigma\ relation, which is due to the rotational broadening of stellar absorption lines in the single-aperture spectra. 
The rotational broadening in single-aperture spectra may systematically increase SVD, particularly for disk galaxies, leading to an artificial trend that black holes are
under-massive (or host galaxies are more massive) than expected from the \msigma\ relation. It is likely that for the majority of pseudo-bulge galaxies,
for which spatially resolved spectroscopic data were not available, the reported SVD based on single aperture in the literature suffer from the rotational broadening,
resulting in an unreliable offset from the \msigma\ relation \citep{Greene+08, Bennert+15}.
Our previous study by \citet{Eun+17} found that a large sample of the hidden type 1 AGN is located slightly below 
the \msigma\ relation with an average offset of $0.12 \pm 0.41$ dex. Again, this conclusion was based on
single-aperture based SVD measurements, suggesting that the effect of rotational broadening introduced the offset.

Various spatially-resolved studies demonstrated that correction for rotational broadening is crucial to properly study the \msigma\ relation. 
 For example, \citet{Bennert+15} used the SVD measurements corrected for galaxy rotation based on spatially-resolved data, to show a sample of 66 local type 1 AGNs follow the  same \msigma\ relations defined by \citet{Woo+13,Woo+15}. 
 \citet{Caglar+20} demonstrated that local luminous AGNs also follow the \msigma\ relation once the SVD measurements are corrected for galaxy rotation.
We also confirm that hidden type 1 AGNs follow the \msigma\ relation albeit with a small sample size, using the properly measured SVDs.


\section{Summary \& Conclusion\label{summary} }

We observed 10 hidden type 1 AGNs using long slit spectrograph to study the spatially resolved kinematics of gas and stars. 
We investigated the nature of AGN gas emission and tested the effect of the rotational broadening in measuring SVDs. 
Our main results are summarized.
 
\begin{itemize}
\item The radial velocity distribution of ionized gas manifested by narrow emission lines show similar trend compared to that of stars,
 indicating a rotation feature. 
\item For seven targets, a broad wing component is detected in the profile of the \oiii\ emission line, indicating that non-gravitational outflows are present at the very center of host galaxies.  
\item We detected a very broad component in the \hb\ line, confirming at least 2 targets (i.e., A09 and A10) are true type 1 AGN. 
\item Using the effective SVD measured from the spatially resolved data, we find that hidden type 1 AGNs follow the \msigma\ relation defined by
more massive inactive galaxies and reverberation-mapped AGN. 
\end{itemize}

\acknowledgments

We thank the referees for various comments, which were useful to improve the clarity of the paper. 
This work was supported by the National Research Foundation of Korea (NRF) grant funded by the Korea government (MEST) (No.  2016R1A2B3011457). 
This work used data that were obtained under the K-GMT Science Program (PID: MMT-2015B-003) funded through Korean GMT Project operated by Korea Astronomy and Space Science Institute.
J.H.W thanks Korea Astronomy and Space Science Institute for its hospitality during a sabbatical visit. 

\end{document}